\documentclass[reprint,aps,pra,superscriptaddress,showpacs,footinbib,longbibliography,10pt]{revtex4-1}
\pdfoutput=1

\usepackage{graphicx}
\usepackage{subfigure}

\usepackage{amsmath}
\usepackage{amssymb}
\usepackage{amsbsy}
\usepackage{wasysym}

\usepackage{bm}
\usepackage{comment}
\usepackage{verbatim}

\usepackage{color}
\usepackage[colorlinks,bookmarks=false,citecolor=darkblue,linkcolor=red,urlcolor=blue]{hyperref}

\definecolor{darkred}{rgb}{0.7,0.0,0.0}
\def\cbl{\color{blue}}
\definecolor{darkblue}{rgb}{0,0.02,0.45}

\definecolor{darkgreen}{rgb}{0.02,0.45,0.0}

\definecolor{violet}{rgb}{0.8,0.2,0.6}

\def\be{\begin{equation}}
\def\ee{\end{equation}}
\def\bea{\begin{eqnarray}}
\def\eea{\end{eqnarray}}

\def\vec{\mathbf}
\def\bs{\boldsymbol}
\def\mc{\mathcal}

\def\cairoold{Bi$_2$Fe$_4$O$_9$}
\def\cairo{Bi$_4$Fe$_5$O$_{13}$F}

\graphicspath{{../}}

\begin{document}

\title{Spin-reorientation transitions in the Cairo pentagonal magnet Bi$_4$Fe$_5$O$_{13}$F}

\author{Alexander A. Tsirlin}
\email{altsirlin@gmail.com}
\affiliation{Experimental Physics VI, Center for Electronic Correlations and Magnetism, University of Augsburg, 86159 Augsburg, Germany}

\author{Ioannis Rousochatzakis}
\email{irousoch@umn.edu}
\affiliation{School of Physics and Astronomy, University of Minnesota, Minneapolis, Minnesota 55455, USA}

\author{Dmitry Filimonov}
\affiliation{Department of Chemistry, Lomonosov Moscow State University, 119991 Moscow, Russia}

\author{Dmitry~Batuk}
\affiliation{EMAT, University of Antwerp, Groenenborgerlaan 171, B-2020 Antwerp, Belgium}

\author{Matthias Frontzek}
\affiliation{Quantum Condensed Matter Division, Oak Ridge National Laboratory, Oak Ridge, TN 37831, USA}

\author{Artem M. Abakumov}
\email{A.Abakumov@skoltech.ru}
\affiliation{Department of Chemistry, Lomonosov Moscow State University, 119991 Moscow, Russia}
\affiliation{Skolkovo Institute of Science and Technology, Nobel str. 3, 143026 Moscow, Russia}

\begin{abstract}
We show that interlayer spins play a dual role in the Cairo pentagonal magnet \cairo, on one hand mediating the three-dimensional (3D) magnetic order and on the other driving spin-reorientation transitions both within and between the planes. 
The corresponding sequence of magnetic orders unraveled by neutron diffraction and M\"ossbauer spectroscopy features two orthogonal magnetic structures described by opposite local vector chiralities, and an intermediate, partly disordered phase with nearly collinear spins. 
A similar collinear phase has been predicted theoretically to be stabilized by quantum fluctuations, but \cairo{} is very far from the relevant parameter regime. 
While the observed in-plane reorientation cannot be explained by any standard frustration mechanism, our {\it ab initio} band-structure calculations reveal strong single-ion anisotropy of the interlayer Fe$^{3+}$ spins that turns out to be instrumental in controlling the local vector chirality and the associated interlayer order. 
\end{abstract}

\maketitle

\section{Introduction}
Frustrated magnets~\cite{HFMBook,Diep,balents2010} host a plethora of remarkable collective phenomena, ranging from topological spin liquids, long-range entanglement and fractionalized excitations~\cite{Anderson73,YanHuseWhite,Shollwock,Han12,Kitaev2006,SavaryBalents,gingras2014}, to emergent electrodynamics and magnetic monopoles~\cite{Castelnovo2008,Henley2010,Rehn2016},
and even to spin-induced ferroelectricity~\cite{mostovoy2007,arima2011,tokura2014}. 
While the majority of geometrically frustrated magnets are based on spin triangles (or tetrahedra in 3D), pentagon-based magnets, which are far more difficult to implement in real materials~\footnote{This is mostly because of the well-known incompatibility of the five-fold symmetry with lattice periodicity.}, are now attracting increasing attention both in theory~\cite{Moessner2001,urumov2002,Raman2005,Jagannathan2011,ralko2011,rojas2012,rousochatzakis2012,pchelkina2013,Nakano2014,Isoda2014} and experiment~\cite{shamir1978,Singh2008,ressouche2009,iliev2010,abakumov2013,Rozova2016,Mohapatra2016}.

\begin{figure*}[!t]
\includegraphics[width=0.9\textwidth,angle=0,clip=true,trim=0 0 0 0]{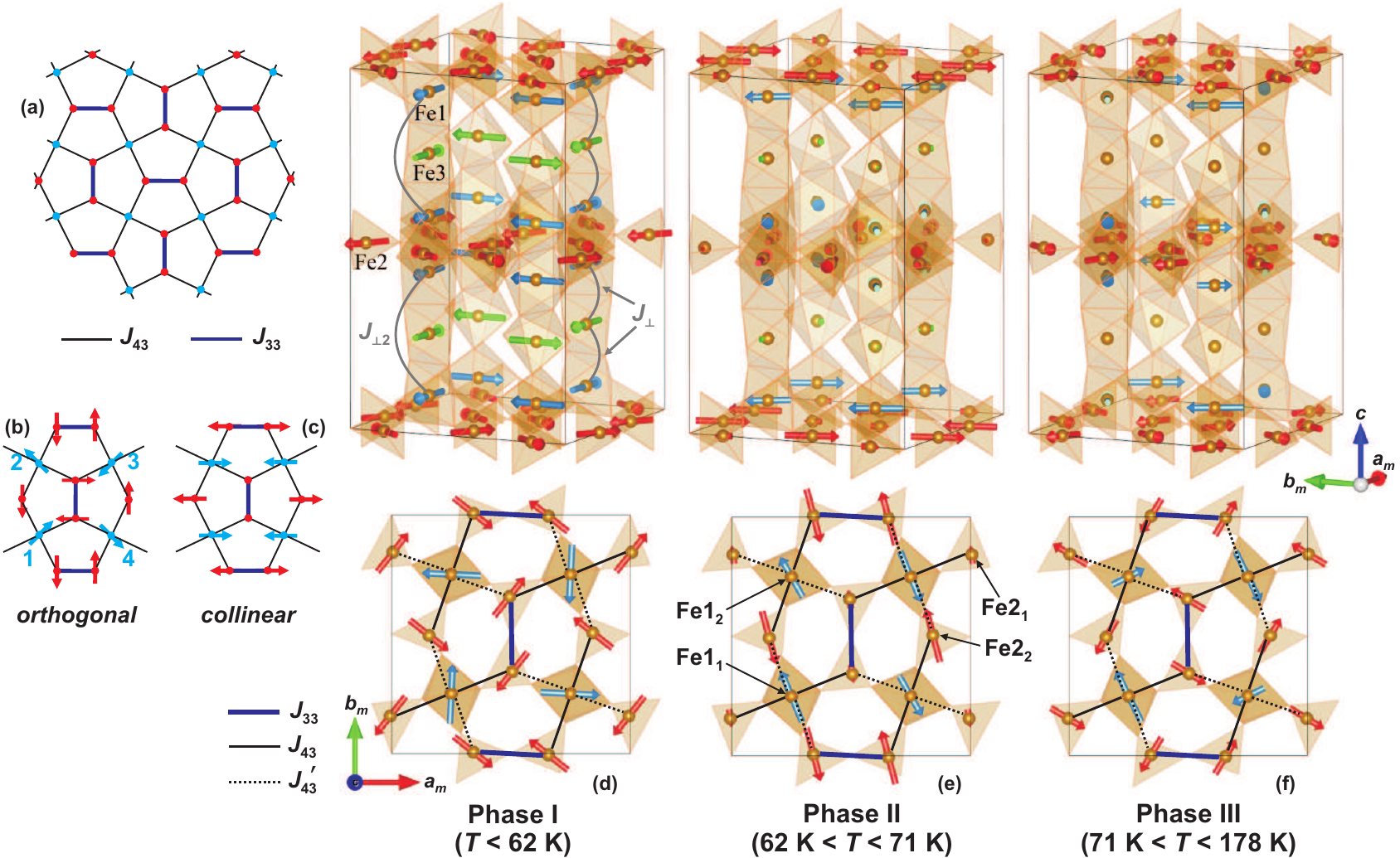}
\caption{\label{fig:structure}
(a) The symmetric Cairo lattice with two exchange couplings, $J_{33}$ and $J_{43}$. (b-c) Orthogonal and collinear phases. Note the zero moment on half of the 3-fold sites in (c). The Fe1 sites 1-4 in (b) provide a measure of local vector chirality as $\bs{\chi}\!=\!\langle \bs{\Gamma}\rangle / |\langle \bs{\Gamma}\rangle|$, where $\bs{\Gamma}\!=\!\vec{S}_1\!\times\!\vec{S}_2\!+\!\vec{S}_2\!\times\!\vec{S}_3\!+\!\vec{S}_3\!\times\!\vec{S}_4\!+\!\vec{S}_4\!\times\!\vec{S}_1$. Note that $\bs{\chi}$ is defined on the plaquette centered by the $\bar 4$ rotoinversion axis (and vertical Fe$2_1$ dimers). The adjacent plaquettes have opposite chirality, but they are centered by the $4_2$ screw axis (and horizontal Fe$2_2$ dimers) and thus distinguished by symmetry. The overall magnetic structure is non-chiral. (d-f) Magnetic structures of \cairo{} in phases I, II, and III, respectively. The two types of the $J_{43}$ couplings are also indicated. The crystal and magnetic structures are visualized using \texttt{VESTA}~\cite{vesta}.
}
\end{figure*}

The main interest so far has been on the Cairo pentagonal lattice, a periodic arrangement of irregular pentagons with two types of sites, one with three-fold and the other with four-fold connectivity (Fig.~\ref{fig:structure}). 
Cairo-based models host various phases of classical and quantum nature~\cite{rousochatzakis2012}, magnetization plateaux~\cite{rousochatzakis2012,ralko2011,Nakano2014,Isoda2014}, and Kosterlitz-Thouless transitions~\cite{ralko2011}. 
By now, there are two main realizations of this lattice, \cairoold{}~\cite{shamir1978,Singh2008,ressouche2009,iliev2010,Mohapatra2016} and \cairo~\cite{abakumov2013}.
A similar pentagonal topology can be also identified in the multiferroics $R$Mn$_2$O$_5$ ($R$ = Bi, Y, or rare-earth)~\cite{Saito1995,Hur2004,blake2005,vecchini2008,harris2008,Sushkov2008,Kim2009,cao2009} that are renown for their complex interplay of commensurate and incommensurate magnetic orders with ferroelectricity.

The symmetric version of the Cairo Heisenberg model has two exchange couplings, $J_{33}$ and $J_{43}$ (Fig.~\ref{fig:structure}\,a), and hosts three phases in the classical, large-$S$ limit~\cite{rousochatzakis2012}: a coplanar orthogonal phase (Fig.~\ref{fig:structure}\,b), a collinear ferrimagnet, and a mixed phase in between. Quantum fluctuations convert the latter to a non-magnetic and possibly spin-nematic phase for $S\!=\!1/2$. Additionally, they introduce another collinear phase for small $J_{43}/J_{33}$~\cite{rousochatzakis2012}. This phase features collinear \textit{antiferromagnetic} order on all four-fold sites and on half of the three-fold sites, with the remaining half being disordered (Fig.~\ref{fig:structure}\,c). 

\cairoold{} and \cairo{} feature dumbbells of the 4-fold-sites. Two Fe1 atoms that comprise a dumbbell lie above and below the Fe2 plane, centered at a nodal point of the pentagonal lattice. Additionally, there are two inequivalent couplings between four-fold and three-fold sites, denoted by $J_{43}$ and $J_{43}'$ (Fig.~\ref{fig:structure}), but the classical phase diagram of this extended model is qualitatively the same (Fig.~\ref{fig:PD}). Furthermore, the calculated interactions (reported here and in~\cite{pchelkina2013}) place the two compounds nearly on the same spot in the phase diagram, and deep inside the orthogonal phase. 
Despite this remarkable similarity, the two Cairo materials show qualitatively different behavior. \cairoold{} orders in the anticipated orthogonal state below 238\,K, but \cairo{}, where Cairo planes are interleaved by an additional layer of Fe3 sites, shows three successive transitions at $T_N\simeq 178$\,K, $T_2\simeq 71$\,K, and $T_1\simeq 62$\,K~\cite{abakumov2013}, with three distinct magnetically ordered states that we refer to as phase I ($T<T_1$), phase II ($T_1<T<T_2$), and phase III ($T_2<T<T_N$). Phase I is orthogonal~\cite{abakumov2013}, whereas the nature of phases II and III is unknown to date.

In the following, we unravel the nature of these phases, and elucidate their origin. Our main findings are:  

i) Phases I and III are both orthogonal and macroscopically non-chiral, but with opposite local vector chiralities, as defined in Fig.~\ref{fig:structure}\,b. Since the two states are degenerate at the level of the isotropic Heisenberg model, anisotropy must play a role. 
 
ii) The intermediate phase II features nearly collinear spins and  drastically reduced moments on half of the Fe2 sites, reminiscent of the quantum collinear phase of \cite{rousochatzakis2012}. This is unexpected given the large, `classical' spin $S\!=\!5/2$ and the fact that we are far from the relevant corner of the phase diagram (Fig.~\ref{fig:PD}).

iii) The onset of phase II upon cooling coincides with a significant growing of the magnetic moments on the interlayer Fe3 sites, which sit between the Cairo planes and normally act to mediate the 3D ordering between the planes, as e.g. in Ref.~\cite{CuP2O6}. 

iv) Spin reorientation within the planes is accompanied by a change of the interlayer order from ferromagnetic (phase I) to antiferromagnetic (phase III).

While these features cannot be explained by any standard frustration mechanism involving purely isotropic (Heisenberg) interactions, our {\it ab initio} band-structure calculations reveal sizable single-ion anisotropy on the interlayer Fe$^{3+}$ spins. 
The Fe3 spins are absent in \cairoold{} (where neighboring planes couple directly to each other), so the emerging physical picture is that the interlayer spins play a vital role for the order within the planes. 
Specifically, the III$\rightarrow$II$\rightarrow$I transitions can be understood as a reorientation of the nominally preferred orthogonal state, from one orientation (phase III) that satisfies the anisotropy on the in-plane spins to another orientation (phase I) that satisfies the anisotropy on the interlayer Fe3 spins, and in between the system must necessarily go through the quasi-collinear phase II. This in-plane spin re-orientation is accompanied by a change in the interlayer order that evolves from antiferromagnetic in phase III to ferromagnetic in phase I.

\section{Magnetic order}
All measurements were performed on single-phase polycrystalline samples of Bi$_4$Fe$_5$O$_{13}$F prepared previously~\cite{abakumov2013}. Neutron diffraction data were collected at the cold neutron powder diffractometer DMC (LNS PSI, Villigen, Switzerland) with the wavelength of 4.5082\,\r A in the $T$ range of $1.5-200$\,K in a He-cryostat. The magnetic structures were refined by the Rietveld method using the \texttt{JANA2006} program~\cite{jana2006}. The symmetry analysis of possible magnetic configurations was carried out in \texttt{ISODISTORT}~\cite{isodistort}.

Phases I--III share the same propagation vector $\mathbf k=(\frac12,\frac12,0)$. The analysis of irreducible representations (irreps) for the $P4_2/mbc$ nuclear structure with this propagation vector yields six irreps. Their corresponding magnetic space groups and allowed magnetic moment components for different positions of the Fe atoms are listed in Table~\ref{tab:irreps}. All these symmetries were tested in the refinement of the $T=1.5$\,K magnetic structure. The magnetic moments were found to be strictly confined to the $ab$ plane. 

The solution was only possible with the $mM_5^-$ irrep and tetragonal magnetic space group $P_C4_2/n$. This magnetic structure can be described within the $\mathbf a_m =\mathbf a-\mathbf b$, $\mathbf b_m=\mathbf a+\mathbf b$, $\mathbf c_m=\mathbf c$ magnetic supercell with five positions (Table~\ref{tab:coordinates}) for the magnetic Fe atoms shown in Fig.~\ref{fig:structure}. We use polar angles $\varphi$ to define orientations of magnetic moments within the plane, $m_a=\mu\cos\varphi$ and $m_b=\mu\sin\varphi$ along the $\mathbf a_m$ and $\mathbf b_m$ directions, respectively.

\begin{table}
\caption{Fractional coordinates of the Fe atoms in the magnetic supercell.\label{tab:coordinates}}
\begin{ruledtabular}
\begin{tabular}{cccc}
        & $x/a$     & $y/b$     & $z/c$  \\\hline
Fe1$_1$ & $\frac14$ & $\frac14$ & 0.0783 \\
Fe1$_2$ & $\frac14$ & $\frac34$ & 0.0783 \\
Fe1$_1$ & 0.0057    & 0.8429    & 0      \\
Fe2$_2$ & 0.3429    & 0.0057    & 0      \\
Fe3     & $\frac14$ & $\frac14$ & $\frac14$ \\
\end{tabular}
\end{ruledtabular}
\end{table}
\begin{table}
\caption{Irreducible representations of $P4_2/mbc$ for $\mathbf k=(\frac12,\frac12,0)$, order parameter directions (OPD), magnetic space groups (mSG), and allowed magnetic moment components for the Fe positions (referred to the unit cell of the nuclear structure).\label{tab:irreps}}
\begin{ruledtabular}
\begin{tabular}{cccccc}
Irrep   		& OPD    	& mSG  			& Fe1			& Fe2			& Fe3  \\\hline
$mM_1^+M_4^+$	& $(a,b)$	& $P_Cccn$		& $\{00z\}$		& $\{xy0\}$ 	& $\{00z\}$ \\
$mM_2^+M_3^+$ 	& $(a,b)$ 	& $P_Cnnn$ 		& $\{00z\}$		& $\{xy0\}$		& none \\
$mM_5^+$ 		& $(a,0)$ 	& $P_C4_2/m$ 	& $\{xy0\}$		& $\{00z\}$	 	& $\{xy0\}$	 \\
		 		& $(a,a)$ 	& $P_Amna$	 	& $\{xy0\}$		& $\{00z\}$	 	& $\{xx0\}$	 \\
		 		& $(a,b)$ 	& $P_a2/m$ 		& $\{xy0\}$		& $\{00z\}$	 	& $\{xy0\}$	 \\
$mM_1^-M_4^-$ 	& $(a,b)$ 	& $P_Cnnm$ 		& $\{00z\}$		& $\{00z\}$		& $\{00z\}$ \\
$mM_2^-M_3^-$ 	& $(a,b)$ 	& $P_Cccm$ 		& $\{00z\}$		& $\{00z\}$	 	& none \\
$mM_5^-$     	& $(a,0)$ 	& $P_C4_2/n$	& $\{xy0\}$		& $\{xy0\}$ 	& $\{xy0\}$ \\
		     	& $(a,a)$ 	& $P_Anna$		& $\{xy0\}$		& $\{xy0\}$ 	& $\{xx0\}$ \\
		     	& $(a,b)$ 	& $P_c2/c$		& $\{xy0\}$		& $\{xy0\}$ 	& $\{xy0\}$ \\
\end{tabular}
\end{ruledtabular}
\end{table}

\subsection{Phase I}
\begin{figure}
\includegraphics[width=8.6cm]{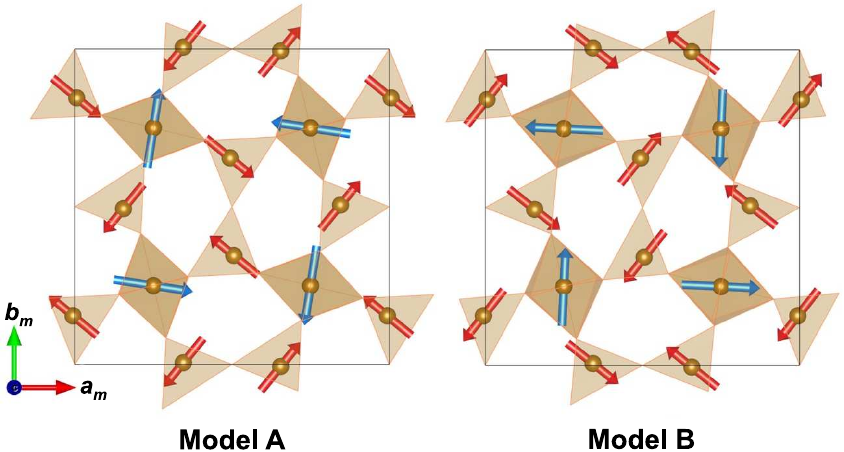}
\caption{\label{fig:models} Distinct magnetic structures A and B at 1.5\,K. The Cairo planes at \textit{z} = 1/2 are shown.}
\end{figure}

At $T=1.5$\,K, two magnetic structures, A and B, are possible, showing very similar arrangement of the magnetic moments in the pentagonal layers (Fig.~\ref{fig:models}). These structures can be transformed into each other by rotating all magnetic moments for about $90^{\circ}$. The main difference between these structures is the orientation of a given Fe1$_1$ moment approximately along $\mathbf a_m$ or $\mathbf b_m$ directions of the magnetic supercell. Although ${\bf a}$ and ${\bf b}$ (and, thus, ${\bf a}_m$ and ${\bf b}_m$) are equivalent in the tetragonal structure, individual Cairo planes lack the tetragonal symmetry. Therefore, anisotropy renders ${\bf a}$ and ${\bf b}$ distinguishable \emph{locally}.

With our earlier HRPT data~\cite{abakumov2013}, models A and B produced virtually the same refinement residuals. Model A was reported as the magnetic structure in phase I~\cite{abakumov2013}, whereas model B was regrettably overlooked in that study. With the DMC data at hand, we can take advantage of the better resolution and sensitivity, and select model B based on the lower refinement residuals, compare $R_{\rm nucl}=0.023$, $R_{\rm mag}=0.027$, and $R_P=0.055$ for the model A and $R_{\rm nucl}=0.021$, $R_{\rm mag}=0.020$, $R_P=0.050$ for the model~B. The choice of model B is further corroborated by the analysis of magnetic anisotropy in Sec.~\ref{sec:anisotropy}.

Phase I (Fig.~\ref{fig:structure}d) is an orthogonal state, with spins on the Fe1$_1$ and Fe1$_2$ sites as well as on the Fe2$_1$ and Fe2$_2$ sites being mutually orthogonal. The interlayer ordering is ferromagnetic, because Fe1 moments of the neighboring Cairo planes interact via Fe3. The antiferromagnetic alignment of Fe1 and Fe3 gives rise to the ferromagnetic arrangement of the Fe1 moments in the adjacent layers. At 1.5\,K, the moments are about 4.0\,$\mu_B$ on the octahedrally coordinated Fe1 and Fe3 sites and 3.3\,$\mu_B$ on the tetrahedrally coordinated Fe2 sites. This difference is due to the stronger Fe--O hybridization for the tetrahedrally coordinated Fe atoms. 

\subsection{Phase II}
\label{sec:phase2}
The propagation vector $\mathbf k=(\frac12,\frac12,0)$ is retained in the entire $T\!=\!1.5-180$\,K temperature range. The refinements were performed assuming that the magnetic structures follow the same irrep and also maintain the order parameter direction and magnetic space group, which would be consistent with weak first-order nature of the transitions at $T_1$ and $T_2$~\cite{abakumov2013}. Satisfactory solutions were found at all temperatures indeed. 

\begin{figure*}
\includegraphics{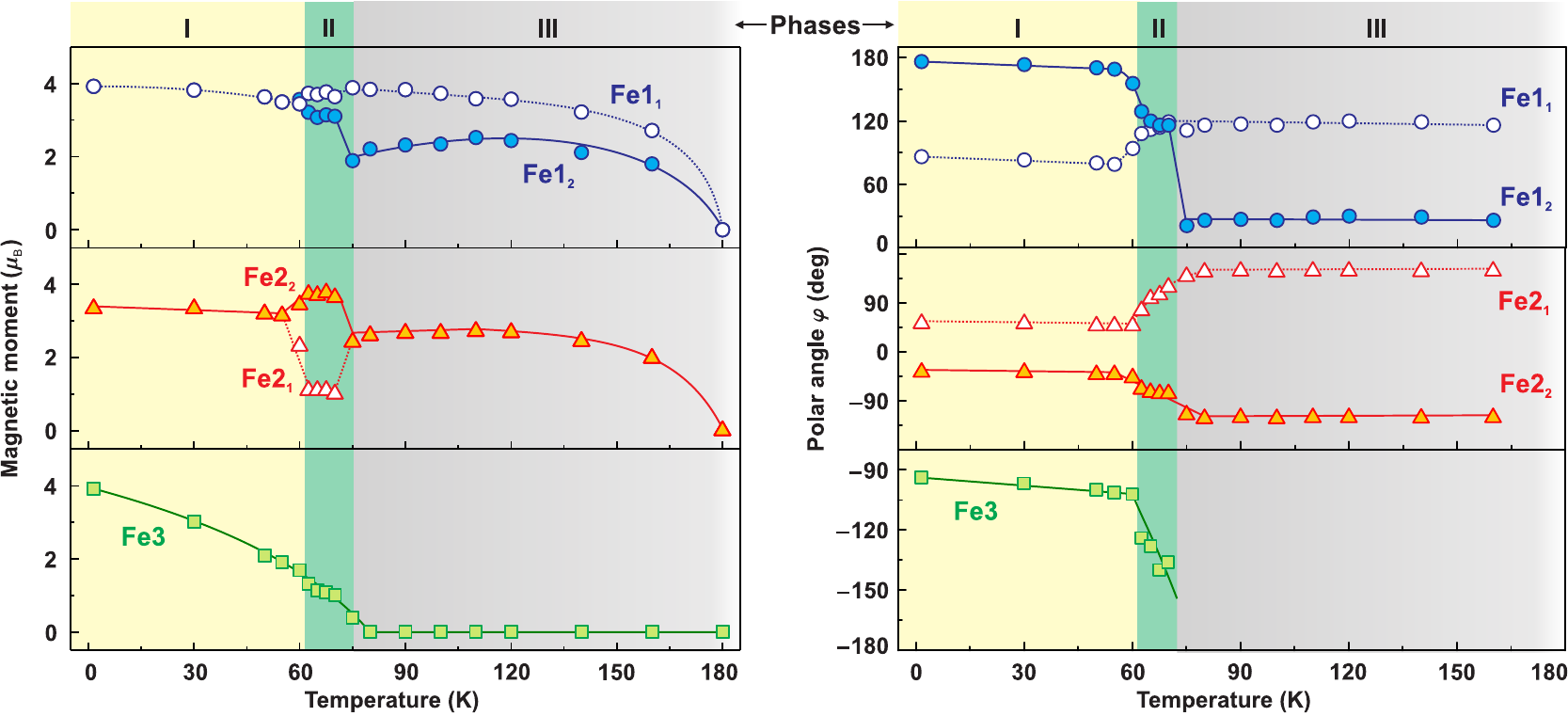}
\caption{\label{fig:moments} Left: temperature dependence of the ordered magnetic moments in Bi$_4$Fe$_5$O$_{13}$F. Right: temperature dependence of the polar angle $\varphi$ showing abrupt rotations of the moments upon the first-order spin-reorientation transitions at $T_1$ and $T_2$. The lines are guide-for-the-eye only.}
\end{figure*}
As $T$ increases toward $T_1$, the Fe1 and Fe2 moments remain roughly unchanged, whereas the moment on Fe3 decreases significantly and drops below 2\,$\mu_B$ at 55\,K (Fig.~\ref{fig:moments}, left). Upon further heating, the magnetic structure changes abruptly entering phase II. All Fe1 sites preserve large moments of $3.2-3.8$\,$\mu_B$, whereas the Fe2 sites split into two groups. The Fe2$_2$ moments increase to 3.8\,$\mu_B$, whereas the Fe2$_1$ moments decrease to about 1.2\,$\mu_B$, only one third of their 1.5\,K value. 

Magnetic moments directions change as well. The in-plane magnetic structure of phase II resembles the quantum collinear phase described in Ref.~\onlinecite{rousochatzakis2012}. Deviations from collinearity are due to the fact that $J_{43}\neq J_{43}'$. In this case, local fields on the Fe2$_1$ site do not cancel, leading to the non-zero ordered moment on Fe2$_1$ and, consequently, to the slight departure of the Fe1$_1$ and Fe1$_2$ moments from the direction of the Fe2$_2$ moment. Therefore, our phase II can be regarded an instance of the collinear phase of the Cairo model for the imperfect realization of this model in \cairo~\footnote{We have also tested a different magnetic structure solution that would impose largely non-collinear spins in the spirit of phase I. To this end, the space group $P_Anna$ was used ($P_Ccnn$ maintaining the crystal axes of $P_C4_2/n$). This solution, however, demonstrated an inferior fit of the magnetic reflections ($R_{\rm mag}=0.030$ at $T=65$\,K) because spin directions for the Fe3 atoms were constrained by symmetry. A better fit might be achievable with the $P_c2/c$ magnetic space group, but it contains 10 independent crystallographic positions for the Fe atoms that renders the solution intractable.}. The reasons behind the formation of this phase are rather unusual, though, and will be discussed in Sec.~\ref{sec:anisotropy}.

The second notable change upon the I$\rightarrow$II transition is the evolution of the interlayer order from ferromagnetic to orthogonal, namely, the adjacent Fe1 moments of the two neighboring Cairo planes, which were parallel in phase~I, become orthogonal in phase II. This is accompanied by the drastic reduction in the ordered moment on Fe3 (Fig.~\ref{fig:moments}).

\subsection{Phase III}
The narrow region of phase II is followed by a broader region of phase III, where in-plane order again becomes orthogonal, whereas the interlayer order returns to collinear, but the overall magnetic structure is quite different from that in phase I. First, the interlayer order is now antiferromagnetic, i.e., adjacent Fe1 moments in the neighboring Cairo planes are antiparallel to each other. Concurrently, the Fe3 moment vanishes. At 100\,K it refined to 0.29(25)\,$\mu_B$, which is insignificant given the experimental error bar. Therefore, we fixed the Fe3 moment to zero throughout the temperature range of phase III.

The differences between the in-plane order of phases I and III can be captured by introducing the local vector chirality $\bs{\chi}$, which we define for the plaquette of four Fe1 spins as explained in the caption of Fig.~\ref{fig:structure}. The magnetic unit cell contains four such plaquettes. Two of them are centered by the $\bar 4$ rotoinversion axis, whereas the other two are centered by the $4_2$ screw axis. The overall $P_C4_2/n$ symmetry requires that adjacent plaquettes have opposite vector chiralities, thus rendering the overall magnetic structure non-chiral. However, each plaquette changes its local vector chirality upon going from phase I to phase III. We find $\bs{\chi}=+\vec{c}$ in phase I and $\bs{\chi}=-\vec{c}$ in phase III for the plaquettes centered by the $\bar 4$ rotoinversion axis (`vertical Fe$2_2$ dimers'), and the other way around for the plaquettes centered by the $4_2$ axis (`horizontal  Fe$2_2$  dimers').

\subsection{M\"ossbauer spectroscopy} 
\begin{figure}
\includegraphics{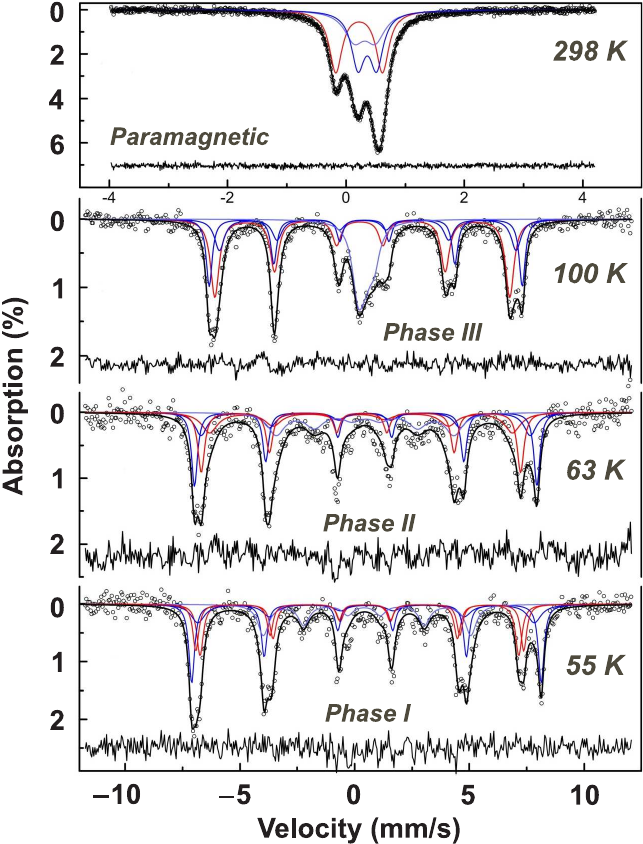}
\caption{\label{fig:moessbauer} M\"ossbauer spectra of Bi$_4$Fe$_5$O$_{13}$F and their fits, as described in the text. For fit parameters, see Table~\ref{tab:moess}.}
\end{figure}

Magnetic structure analysis was supported by M\"ossbauer spectroscopy measurements. The $^{57}$Fe M\"ossbauer spectra (Fig.~\ref{fig:moessbauer}) were recorded in the temperature range $55-300$\,K in a transmission mode with a $^{57}$Co/Rh $\gamma$-ray source using a constant acceleration spectrometer MS1104. At room temperature, the spectrum can be decomposed into 3 doublets with the nearly 40:40:20 ratio of the intensities corresponding to the Fe1, Fe2, and Fe3 positions, respectively (Table~\ref{tab:moess}). 

Upon cooling below $T_N$, the spectra reveal an additional splitting indicative of the magnetic ordering. However, the spectrum at 100\,K, in phase III, could not be accounted for by a combination of regular sextets. We find that about 20\% of the spectral intensity corresponds to an unresolved sextet with a very weak hyperfine splitting. This signal arises from Fe3 sites that, according to the neutron data, feature negligible ordered moment above $T_2$. Below $T_2$, the Fe3 moments increase and a well-resolved sextet develops. 

\begin{table}
\caption{\label{tab:moess}
Parameters of the M\"ossbauer spectra for Bi$_4$Fe$_5$O$_{13}$F. $I$ stands for the fraction of the spectral intensity, $\delta$ is the isomer shift, $\Delta EQ$ is the quadrupolar splitting, $\Gamma$ is the linewidth, and $H$ is the hyperfine field.
}
\begin{ruledtabular}
\begin{tabular}{lccrcc}
    & $I$  & $\delta$ & $\Delta EQ$ & $\Gamma$ & $H$   \\
	  & (\%) & (mm/s)   & (mm/s)      & (mm/s)   & (T) \\\hline
	  & \multicolumn{5}{c}{$T=300$\,K}                   \\
Fe1 & 35   & 0.37     & 0.31        & 0.27     & --    \\
Fe2 & 38   & 0.23     & 0.78        & 0.26     & --    \\
Fe3 & 27   & 0.33     & 0.36        & 0.40     & --    \smallskip\\
    & \multicolumn{5}{c}{$T=100$\,K}                   \\
Fe1$_1$ & 23 & 0.45   & 0.06        & 0.32     & 42.4 \\
Fe1$_2$ & 17 & 0.46   & 0.21        & 0.52     & 40.2 \\
Fe2     & 38 & 0.28   & 0.11        & 0.45     & 39.9 \\
Fe3     & 22 & 0.41   & 0.18        & 0.60     & 22  \smallskip\\
    & \multicolumn{5}{c}{$T=63$\,K}                   \\
Fe1$_1$ & 24 & 0.46   & 0.04        & 0.31     & 46.4 \\
Fe1$_2$ & 16 & 0.46   & 0.03        & 0.67     & 44.6 \\
Fe2$_1$ & 18 & 0.29   & 0.00        & 0.87     & 41.0 \\
Fe2$_2$ & 22 & 0.29   & $-0.04$     & 0.35     & 43.2 \\
Fe3     & 21 & 0.47   & $-0.05$     & 0.93     & 24.1 \smallskip\\
    & \multicolumn{5}{c}{$T=55$\,K}                   \\
Fe1$_1$ & 27 & 0.50   & 0.05        & 0.30     & 47.6 \\
Fe1$_2$ & 15 & 0.52   & $-0.03$     & 0.73     & 45.8 \\
Fe2$_1$ & 17 & 0.34   & $-0.23$     & 0.31     & 44.6 \\
Fe2$_2$ & 20 & 0.34   & $-0.26$     & 0.34     & 43.4 \\
Fe3     & 21 & 0.48   & 0.13        & 0.58     & 28.3 \\
\end{tabular}
\end{ruledtabular}
\end{table}

Below $T_2$, the spectrum is decomposed into 5 sextets (Table~\ref{tab:moess}). The reduced ordered moment on Fe2$_1$ manifests itself in the largely broadened signal. At 55\,K, in phase I, the sextets of Fe2$_1$ and Fe3 become more narrow suggesting the formation of large magnetic moments on all Fe sites, which is again in agreement with the magnetic structures shown in Fig.~\ref{fig:structure}. 

\section{Magnetic model}
\subsection{Isotropic exchange couplings}
The isotropic exchange couplings in \cairo\ were previously reported in Ref.~[\onlinecite{abakumov2013}]. However, using these values in numerical simulations of the magnetic susceptibility, we arrived at the too high $T_N\!=\!250$\,K compared to the experimental value of 178\,K. Therefore, we revised the microscopic magnetic model using extensive density-functional (DFT) band-structure calculations~\cite{pbe96} performed in the \texttt{FPLO}~\cite{fplo} and \texttt{VASP}~\cite{vasp1,vasp2} codes. Total energies of collinear spin configurations were mapped onto the spin Hamiltonian, and exchange couplings were determined~\cite{xiang2013}. The accuracy of this approach was improved by choosing different values of the on-site Coulomb repulsion for the octahedrally coordinated Fe1 and Fe3 sites and for the tetrahedrally coordinated Fe2 sites, following different oxygen coordination and, therefore, different screening~\cite{SM}. 

\begin{figure}
\includegraphics{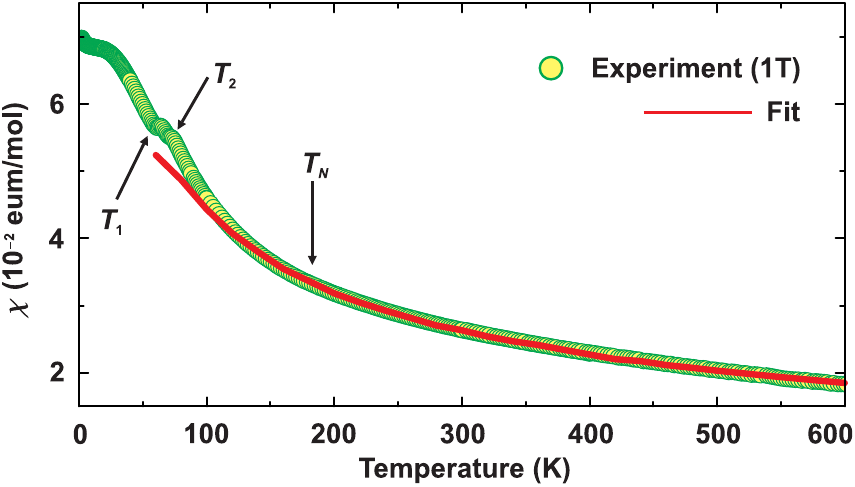}
\caption{\label{fig:fit} Fit of the magnetic susceptibility of Bi$_4$Fe$_5$O$_{13}$F~\cite{abakumov2013} using revised exchange parameters reported in this work.}
\end{figure}

DFT-based exchange couplings were refined by Monte-Carlo simulation of the magnetic susceptibility. The optimized set of exchange parameters was obtained, with $J_{33}=116$\,K, $J_{43}=38$\,K, $J_{43}'=57$\,K, and the Fe1--Fe3 inter-plane interaction $J_{\perp}=8$\,K (Fig.~\ref{fig:structure}). Additionally, a weak second-neighbor interlayer Fe1--Fe1 interaction $J_{\perp 2}=2$\,K is present. 
This set of parameters reproduces the susceptibility down to 120\,K (Fig.~\ref{fig:fit}) and predicts $T_N\simeq 180$\,K in perfect agreement with the experiment.

\subsection{Magnetic anisotropy}
\label{sec:anisotropy}
Several anisotropic terms may occur in \cairo. Their calculation follows the same procedure~\cite{xiang2013} with the only exception that orthogonal spin configurations are used~\cite{SM}. Symmetric exchange anisotropy corresponds to energies well below 0.1\,K per Fe atom and is thus negligible. The antisymmetric exchange anisotropy, Dzyaloshinsky-Moriya (DM) interactions allowed on the $J_{43}$, $J_{43}'$, and $J_{\perp}$ bonds, are stronger, up to about 5\,K per Fe atom, but their effect on the magnetic structure largely cancels out, because in the orthogonal structures of phases I and III the adjacent $J_{43}$ ($J_{43}'$) bonds feature same directions of the DM vectors, yet opposite spin rotations. 

Single-ion anisotropy terms are believed to be small in Fe$^{3+}$ compounds due to the $d^5$ nature of the magnetic ion. Unexpectedly, we find that these terms are in fact non-negligible and central to the physics of \cairo. Single-ion anisotropy is obtained by fixing spins along a given direction and rotating the reference spin in the plane perpendicular to this direction~\cite{SM}. Angular dependence of the energy, $E(\varphi)$, directly measures the single-ion anisotropy of the reference spin.

\begin{figure}
\includegraphics{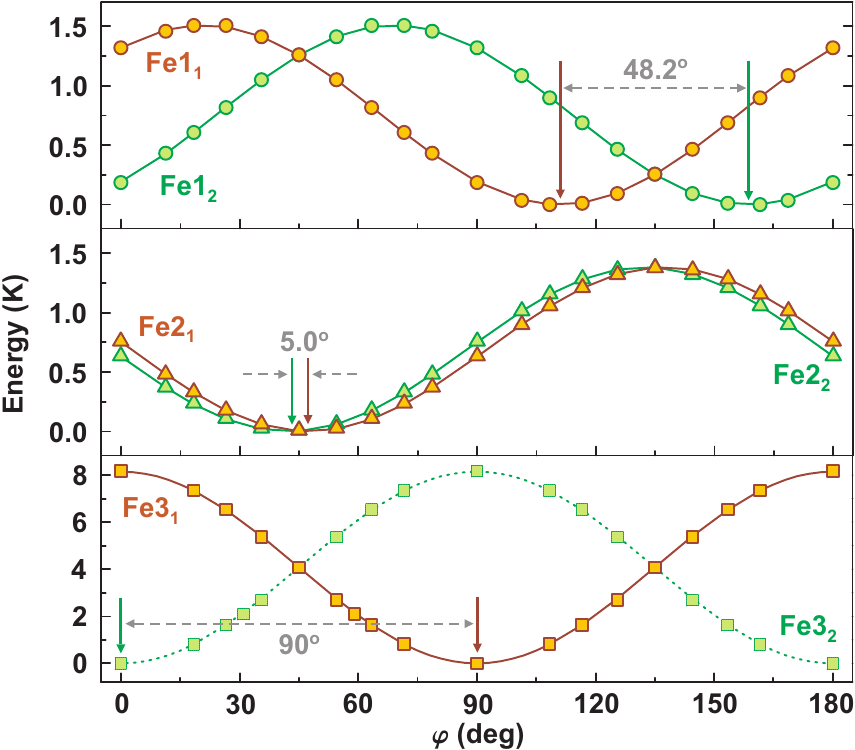}
\caption{\label{fig:SIA}
In-plane single-ion anisotropy energies for the Fe1, Fe2, and Fe3 sites. The arrows denote preferred spin directions. Only the Fe3 sites are compatible with the orthogonal structure, because their preferred directions are at $90^{\circ}$ to each other. For the Fe1 and Fe2 sites, the angle between preferred directions of the neighboring sites is largely different from $90^{\circ}$. Note that we use Fe3$_1$ and Fe3$_2$ for those Fe3 sites that are coupled to Fe1$_1$ and Fe1$_2$, respectively. The angle $\varphi$ is measured between the magnetic moment and the $\mathbf a_m$-axis, and all curves are periodic, $E(\varphi+180^{\circ})=E({\varphi})$.
}
\end{figure}
Fig.~\ref{fig:SIA} shows single-ion anisotropies for different Fe sites. The single-ion anisotropy of Fe3 is much stronger than those of Fe1 and Fe2. This can be attributed to a large distortion of the Fe3O$_6$ octahedra~\cite{abakumov2013}. The positions of the energy minima are compatible with the symmetry of the crystal structure, where mirror planes require that the $E(\varphi)$ curves are symmetric with respect to $\varphi=45^{\circ}$ and $135^{\circ}$. 

An immediate effect of the single-ion anisotropy terms is the selection between states A and B that can form in phase I. Here, the predominant single-ion anisotropy of Fe3 favors either $\mathbf a_m$ or $\mathbf b_m$ direction of a given Fe3 atom. The Fe1 spins choose same directions because of the isotropic coupling $J_{\perp}$. Therefore, the Fe1$_1$ spins should point along the $\mathbf b_m$ axis, whereas the Fe1$_2$ spins should point along the $\mathbf a_m$ axis, as seen in the state B that is pinpointed by our neutron data.

The single-ion anisotropy also plays central role for the selection of both in-plane and interlayer order in \cairo, as we explain below.

\subsection{In-plane order}
\begin{figure}[t]
\includegraphics[width=8.3cm]{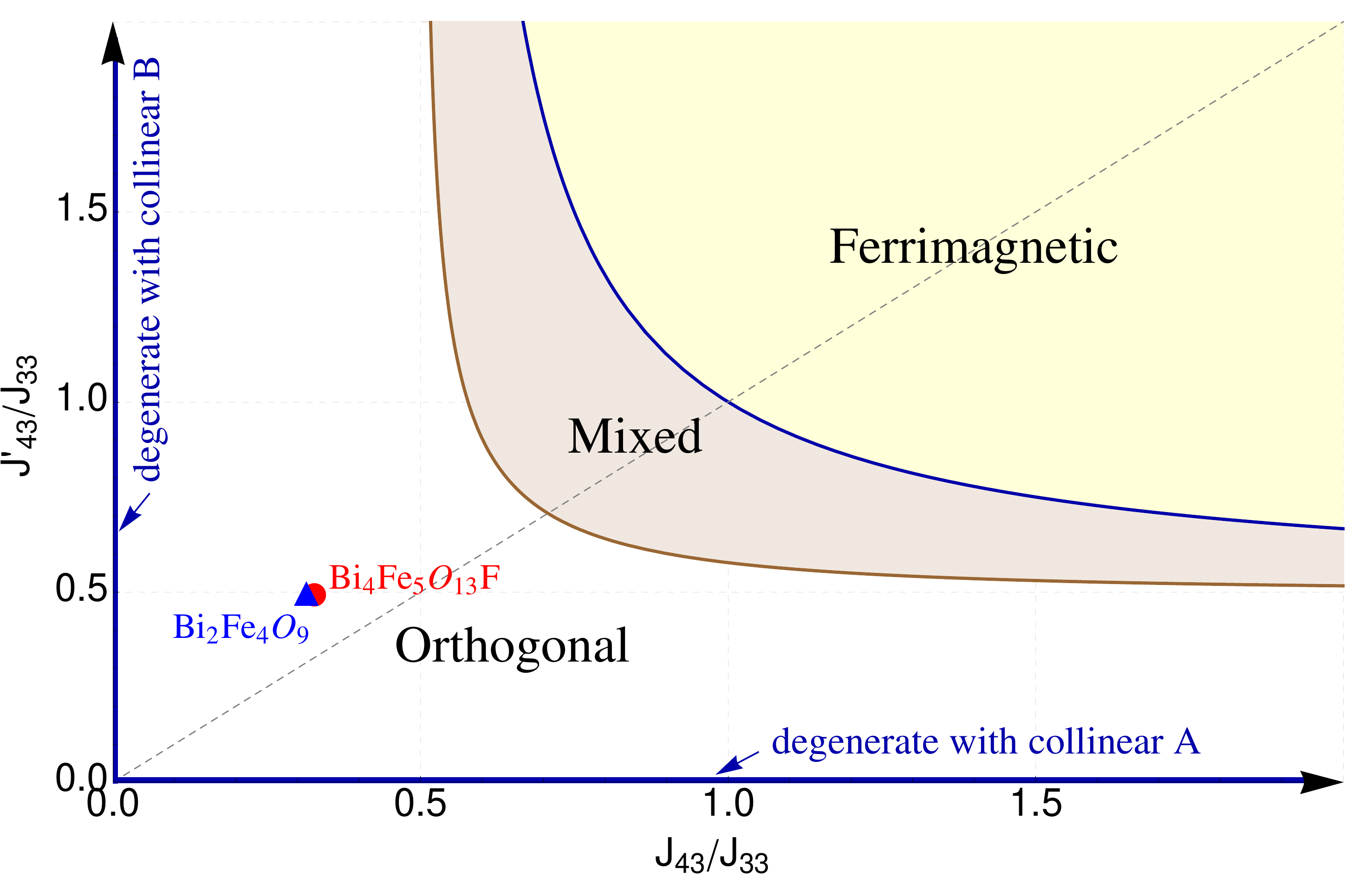}
\caption{\label{fig:PD}
Classical phase diagram of the isotropic $J_{43}-J_{43}'-J_{33}$ model, with two Fe1 spins on each four-fold site of the lattice, see text. The two available compounds are shown by the filled blue triangle (based on the parameters of \cite{pchelkina2013}) and the filled red dot (based on the parameters given above).
}
\end{figure}

The classical phase diagram of the $J_{43}-J_{43}'-J_{33}$ model that describes the in-plane isotropic interactions is shown in Fig.~\ref{fig:PD}. 
This model takes into account the two inequivalent Fe1--Fe2 couplings, $J_{43}$ and $J_{43}'$, and the fact that there are two Fe1 spins on each four-fold site of the lattice.
The phase diagram has been obtained using Lyons and Kaplan's~\cite{LyonsKaplan,*Kaplan} generalization of the Luttinger-Tisza method~\cite{LT}, see Ref.~\cite{SM} for technical details.
The phase diagram contains three main phases, the coplanar orthogonal phase, the collinear ferrimagnetic phase, and a mixed phase in between, which is non-coplanar. In the latter phase, the projections of the spins along an axis yields the ferrimagnetic configuration while the projections in the plane orthogonal to that axis yields the orthogonal configuration, see also Ref.~\onlinecite{rousochatzakis2012}. The relative projections interpolate between zero and one as we go across the two boundaries of this phase.

The special lines $J_{43}'\!=\!0$ and $J_{43}\!=\!0$ correspond to decoupled chains. Along these lines, the orthogonal phase becomes degenerate with infinite other ground states, including the so-called collinear phases A and B discussed in~\cite{SM}. The partially disordered collinear phase of \cite{rousochatzakis2012}, reminiscent of phase II of Fig.~\ref{fig:structure}\,(e), is stabilized by quantum fluctuations in the corner around $J_{43}\!=\!J_{43}'\!=\!0$. The line $J_{43}\!=\!J_{43}'$ maps to the model of \cite{rousochatzakis2012} by rescaling $J_{43}\!\to\!J_{43}/2$ and $J_{43}'\!\to\!J_{43}'/2$. 

Based on the above {\it ab initio} values, \cairo\ sits deep inside the orthogonal phase (filled red dot in Fig.~\ref{fig:PD}), and far away from the corner $J_{43}\!=\!J_{43}'\!=\!0$. Interestingly, the second Cairo magnet, \cairoold, sits almost on the same spot of the phase diagram (filled blue triangle in Fig.~\ref{fig:PD}), according to  the {\it ab initio} parameters of \cite{pchelkina2013}.

The interlayer Fe1--Fe3 coupling is much weaker than the couplings within the plane. Therefore, one expects that the Fe3 spins are more sensitive to thermal fluctuations and decrease much faster than the spins on Fe1 and Fe2~\cite{SM}, in agreement with Fig.~\ref{fig:moments}. On the other hand, the formation of phase II can not be anticipated, because the system is far away from any collinear phase at zero temperature (Fig.~\ref{fig:PD}). There is a large energy barrier against any thermally-driven stabilization of collinearity at the level of the isotropic model. The scenario of quantum fluctuations driving the collinear phase is unlikely as well. The onset of the collinear phase is roughly taking place when the spin length correction $\delta S$ from quadratic spin waves approaches the full value $S\!=\!5/2$. According to Fig.~4 of Ref.~\cite{rousochatzakis2012}, the collinear phase for $S=5/2$ (if any) onsets way below $J_{43}/J_{33}\!=\!0.1$, and this number should be further divided by two, because here we have two Fe1 sites at each four-fold site. 
So the formation of phase II requires the presence of anisotropy. 

\begin{figure*}
\includegraphics[scale=0.78]{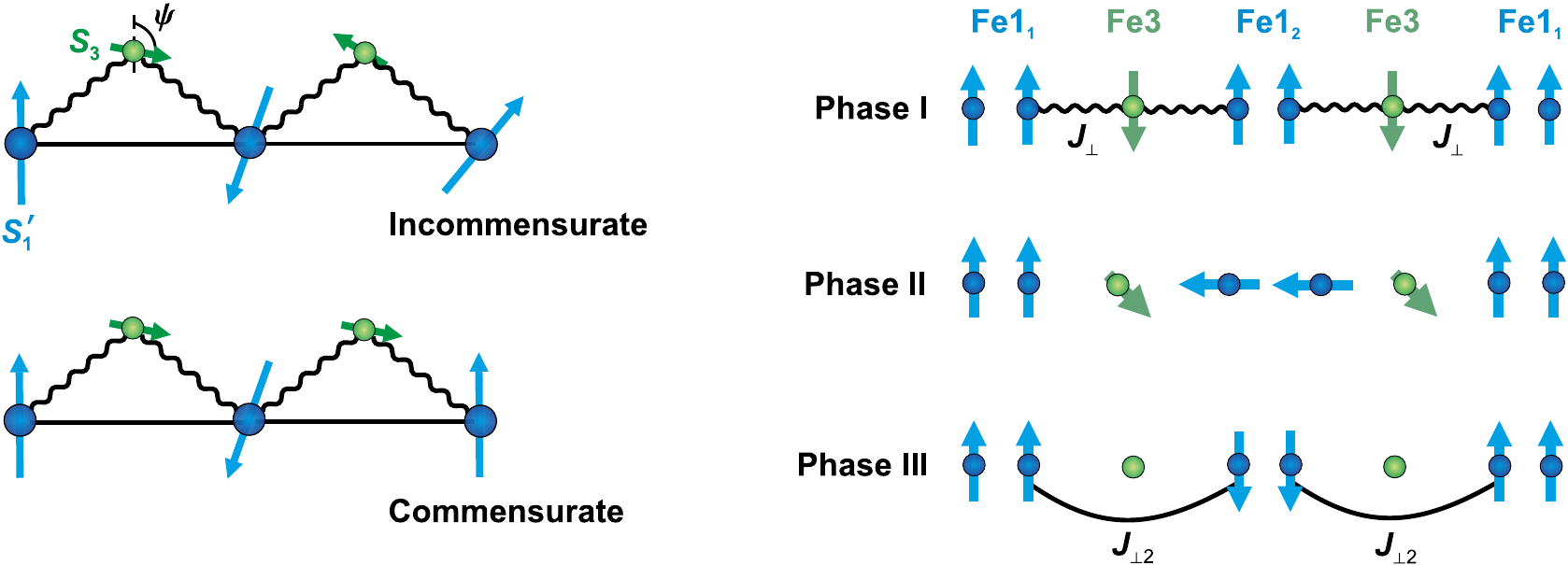}
\caption{\label{fig:sawtooth}
Left panel: sawtooth-chain model with the couplings $J_{\perp}$ and $J_{\perp2}$ and two ordered states, incommensurate and commensurate, which are degenerate on the classical level. Right panel: the interlayer order in \cairo{}. 
}
\end{figure*}
The anisotropy of Fe3 is more than 5 times stronger than that of Fe1 and Fe2. Therefore, at low temperatures, in phase I, Fe3 with its preferred directions at $\varphi=0^{\circ}$ and $90^{\circ}$ puts the Fe1 moments along $\mathbf a_m$ and $\mathbf b_m$. It does not choose the flavor of vector chiral order per se, but the anisotropy of Fe2 selects $\bs{\chi}\!=\!+\vec{c}$ in phase I, as confirmed by a direct energy minimization.

Above $T_2$, in phase III, the preferred direction of Fe3 plays no role, and the Fe1 and Fe2 moments are left to form an orthogonal configuration, even though their preferred directions are not compatible with such a structure. For example, the preferred directions of Fe1$_1$ and Fe1$_2$ differ by $48.2^{\circ}$ only, whereas Fe2$_1$ and Fe2$_2$ moments prefer to be nearly collinear. However, deviations from the orthogonal structure cost a lot of exchange energy, because the exchange couplings are at least two orders of magnitude stronger than the anisotropy. A clear fingerprint of this competition between the orthogonal state and individual single-ion anisotropies is the large and unexpected difference in the magnetic moments of Fe1$_1$ and Fe1$_2$ in phase III. Indeed, the Fe1$_1$ moment is larger, because it is close to the preferred direction (the departure from the preferred direction is $\Delta\varphi=5^{\circ}$ at 100\,K). On the other hand, the moment on Fe1$_2$ is far away from its preferred direction ($\Delta\varphi=47^{\circ}$) and thus 30\,\% smaller.

A side effect of these energy considerations is that vector chiral order changes from $\bs{\chi}\!=\!+\vec{c}$ in phase I to $\bs{\chi}\!=\!-\vec{c}$ in phase III. The continuous transformation between these two phases necessitates the intermediate quasi-collinear phase II that exists in a narrow temperature range only.

\subsection{Interlayer order}
Let us now turn to the interlayer order that can be described by an effective 1D model of the --Fe1$_1$--Fe3--Fe1$_2$--Fe1$_2$--Fe3--Fe1$_1$-- chain. It is essentially a ferrimagnetic chain, where $S_1'=5$ stands for the Fe1 dumbbell and $S_3=\frac52$ stands for the Fe3 atom. The nearest-neighbor Fe1--Fe3 coupling $J_{\perp}$ is augmented by the next-nearest-neighbor Fe1--Fe1 coupling $J_{\perp2}$ resulting in a sawtooth-chain geometry (Fig.~\ref{fig:sawtooth}). Classical energy minimization for such a model gives rise to a non-collinear state with the angle $\psi$ between the neighboring spins given by
\begin{equation}
 \cos\psi=-\frac{J_{\perp}\,S_3}{2J_{\perp2}\,S_1'}.
\end{equation}
Two $\psi$ rotations may be followed by another two $\psi$ rotations or by two $-\psi$ rotations. Therefore, there is infinite classical degeneracy in the $J_{\perp}-J_{\perp2}$ sawtooth-chain model, because any sequence of pairwise $\psi$ and $-\psi$ rotations can occur. This situation is remarkably similar to kagome francisites~\cite{rousochatzakis2015}, where same physics is observed on a 2D lattice, and the ground state is chosen (already on the classical level) by anisotropic terms in the spin Hamiltonian. 

In the case of Bi$_4$Fe$_5$O$_{13}$F, $J_{\perp}/J_{\perp2}\simeq 4$ and $S_3/S_1'\simeq\frac12$ produce ferrimagnetic order along the $c$ direction at low temperatures (phase I, Fig.~\ref{fig:sawtooth}). At higher temperatures, $S_3$ decreases, and the $S_3/S_1'$ ratio decreases as well. Therefore, $2\psi$ departs from $360^{\circ}$ and eventually reaches $180^{\circ}$ in phase III, where $S_3=0$, and the interlayer order is antiferromagnetic (Fig.~\ref{fig:sawtooth}). Phase II is in the intermediate regime with $2\psi=270^{\circ}$ (Fig.~\ref{fig:sawtooth}), i.e., the interlayer order is orthogonal with the $90^{\circ}$ configuration between the Fe1 moments in the adjacent planes.

Ground state selection requires anisotropic terms in the spin Hamiltonian. The $J_{\perp}-J_{\perp2}$ sawtooth-chain model makes no difference between the commensurate (``canted'') state, where two $\psi$ rotations are followed by two $-\psi$ rotations, and the incommensurate (helical) state, where only $\psi$ rotations occur (Fig.~\ref{fig:sawtooth}, left). This degeneracy is lifted by anisotropy terms. 

In Bi$_4$Fe$_5$O$_{13}$F, single-ion anisotropy of Fe1 and Fe3 is at play. This anisotropy favors same spin directions on all Fe3 atoms and, respectively, on all Fe1 atoms from every second Cairo plane. Therefore, the commensurate order along the ${\bf c}$ direction is stabilized. It is worth noting that the DM coupling on the $J_{\perp}$ bonds would have an opposite effect and favor the incommensurate helical order, but such a coupling is smaller than the single-ion anisotropy terms providing the energy of 0.4\,K per Fe atom only~\footnote{For the sake of comparison with Fig.~\ref{fig:SIA}, we provide the overall coupling energy that is \textit{not} normalized to $S=\frac52$.}. Therefore, the single-ion anisotropy is crucial not only for the in-plane order, but also for the commensurate nature of the order between the Cairo planes.

\section{Discussion and Conclusions} 
The main picture emerging from the experimental data presented here is that the interlayer Fe3 spins in \cairo{} play a dual role, on one hand mediating the 3D ordering and on the other driving a reorientation of the order both within and between the planes. 
While details of this transition require further dedicated theoretical work, on the experimental side the effect of the Fe3 spins is crucial for the design of new Cairo-lattice magnets, because interlayer magnetic sites, which are often introduced for the sake of stabilizing the crystal structure~\cite{cumby2016}, are not innocent and in fact play decisive role for the magnetic order. 

The sequence of transitions in \cairo{} is reminiscent of the consecutive spin re-orientations in $R$Mn$_2$O$_5$, where an intermediate collinear phase separates two non-collinear states. However, this collinear phase~\cite{blake2005} is different from our phase II, because it does not show the characteristic reduction in the ordered moment on half of the Fe2 sites. Instead, it may be related to the collinear phases A and B of Fig.~\ref{fig:PD}. 

More generally, we show that magnetic order in the pentagonal geometry is largely influenced by even weak anisotropy terms. Despite structural similarities, different systems may eventually show very different types of magnetic order depending on the transition-metal ion. In the case of RMn$_2$O$_5$, the $d^4$ Mn$^{3+}$ ion is known to be more anisotropic than the $d^5$ Fe$^{3+}$~\footnote{The typical single-ion anisotropy term for octahedrally coordinated Mn$^{3+}$ is $2-3$\,K~\cite{moussa1996,kida1973} yielding the energy difference on the order of 10\,K upon spin rotation, the energy scale similar to that of Fe3 in Fig.~\ref{fig:SIA}.}, and no direct analogies between Fe-based Cairo magnets and multiferroic RMn$_2$O$_5$ manganites may occur.

Phase II of \cairo\ is quite unusual on its own. On one hand, it strongly resembles quantum collinear phase of the Cairo model~\cite{rousochatzakis2015}. As explained in Sec.~\ref{sec:phase2}, deviations from the collinearity in this phase are due to the imperfect nature of the Cairo lattice ($J_{43}\neq J_{43}'$). On the other hand, phase~II in \cairo\ is not stabilized by quantum fluctuations and originates from competing single-ion anisotropies. Despite this different origin, phase II involves significant amount of fluctuations reflected in the low ordered moment on Fe2$_1$. It would be interesting to explore whether properties of phase II and especially its magnetic excitations are similar to those of the quantum collinear phase established in Ref.~\onlinecite{rousochatzakis2015}. Moreover, variable magnetic structures of \cairo\ may have effect on its hitherto unknown dielectric behavior.

\begin{acknowledgments}
We are grateful to Prof. J.-M. Perez-Mato and Dr. Dmitry Khalyavin for valuable discussions on the magnetic structures and symmetries. DF and AA are grateful to the Russian Science Foundation (grant 14-13-00680) for support. AT was supported by the Federal Ministry for Education and Research through the Sofja Kovalevskaya Award of Alexander von Humboldt Foundation. This work is based on experiments performed at the Swiss spallation neutron source SINQ, Paul Scherrer Institut, Villigen, Switzerland.
\end{acknowledgments}


\begin{thebibliography}{68}%
\makeatletter
\providecommand \@ifxundefined [1]{%
 \@ifx{#1\undefined}
}%
\providecommand \@ifnum [1]{%
 \ifnum #1\expandafter \@firstoftwo
 \else \expandafter \@secondoftwo
 \fi
}%
\providecommand \@ifx [1]{%
 \ifx #1\expandafter \@firstoftwo
 \else \expandafter \@secondoftwo
 \fi
}%
\providecommand \natexlab [1]{#1}%
\providecommand \enquote  [1]{``#1''}%
\providecommand \bibnamefont  [1]{#1}%
\providecommand \bibfnamefont [1]{#1}%
\providecommand \citenamefont [1]{#1}%
\providecommand \href@noop [0]{\@secondoftwo}%
\providecommand \href [0]{\begingroup \@sanitize@url \@href}%
\providecommand \@href[1]{\@@startlink{#1}\@@href}%
\providecommand \@@href[1]{\endgroup#1\@@endlink}%
\providecommand \@sanitize@url [0]{\catcode `\\12\catcode `\$12\catcode
  `\&12\catcode `\#12\catcode `\^12\catcode `\_12\catcode `\%12\relax}%
\providecommand \@@startlink[1]{}%
\providecommand \@@endlink[0]{}%
\providecommand \url  [0]{\begingroup\@sanitize@url \@url }%
\providecommand \@url [1]{\endgroup\@href {#1}{\urlprefix }}%
\providecommand \urlprefix  [0]{URL }%
\providecommand \Eprint [0]{\href }%
\providecommand \doibase [0]{http://dx.doi.org/}%
\providecommand \selectlanguage [0]{\@gobble}%
\providecommand \bibinfo  [0]{\@secondoftwo}%
\providecommand \bibfield  [0]{\@secondoftwo}%
\providecommand \translation [1]{[#1]}%
\providecommand \BibitemOpen [0]{}%
\providecommand \bibitemStop [0]{}%
\providecommand \bibitemNoStop [0]{.\EOS\space}%
\providecommand \EOS [0]{\spacefactor3000\relax}%
\providecommand \BibitemShut  [1]{\csname bibitem#1\endcsname}%
\let\auto@bib@innerbib\@empty
\bibitem [{HFM(2011)}]{HFMBook}%
  \BibitemOpen
  \href@noop {} {\emph {\bibinfo {title} {Introduction to Frustrated Magnetism:
  Materials, Experiments, Theory}}}\ (\bibinfo  {publisher} {Springer Series in
  Solid-State Sciences},\ \bibinfo {address} {Berlin},\ \bibinfo {year}
  {2011})\BibitemShut {NoStop}%
\bibitem [{Die(2013)}]{Diep}%
  \BibitemOpen
  \href@noop {} {\emph {\bibinfo {title} {Frustrated Spin Systems}}},\ \bibinfo
  {edition} {2nd}\ ed.\ (\bibinfo  {publisher} {World Scientific},\ \bibinfo
  {year} {2013})\BibitemShut {NoStop}%
\bibitem [{\citenamefont {Balents}(2010)}]{balents2010}%
  \BibitemOpen
  \bibfield  {author} {\bibinfo {author} {\bibfnamefont {L.}~\bibnamefont
  {Balents}},\ }\bibfield  {title} {\enquote {\bibinfo {title} {Spin liquids in
  frustrated magnets},}\ }\href@noop {} {\bibfield  {journal} {\bibinfo
  {journal} {Nature}\ }\textbf {\bibinfo {volume} {464}},\ \bibinfo {pages}
  {199--208} (\bibinfo {year} {2010})}\BibitemShut {NoStop}%
\bibitem [{\citenamefont {Anderson}(1973)}]{Anderson73}%
  \BibitemOpen
  \bibfield  {author} {\bibinfo {author} {\bibfnamefont {P.W.}\ \bibnamefont
  {Anderson}},\ }\bibfield  {title} {\enquote {\bibinfo {title} {Resonating
  valence bonds: A new kind of insulator?}}\ }\href
  {http://www.sciencedirect.com/science/article/pii/0025540873901670}
  {\bibfield  {journal} {\bibinfo  {journal} {Mat. Res. Bull}\ }\textbf
  {\bibinfo {volume} {8}},\ \bibinfo {pages} {153 -- 160} (\bibinfo {year}
  {1973})}\BibitemShut {NoStop}%
\bibitem [{\citenamefont {Yan}\ \emph {et~al.}(2011)\citenamefont {Yan},
  \citenamefont {Huse},\ and\ \citenamefont {White}}]{YanHuseWhite}%
  \BibitemOpen
  \bibfield  {author} {\bibinfo {author} {\bibfnamefont {S.}~\bibnamefont
  {Yan}}, \bibinfo {author} {\bibfnamefont {D.~A.}\ \bibnamefont {Huse}}, \
  and\ \bibinfo {author} {\bibfnamefont {S.~R.}\ \bibnamefont {White}},\
  }\bibfield  {title} {\enquote {\bibinfo {title} {Spin-liquid ground state of
  the $s=1/2$ kagome {Heisenberg} antiferromagnet},}\ }\href {\doibase
  10.1126/science.1201080} {\bibfield  {journal} {\bibinfo  {journal}
  {Science}\ }\textbf {\bibinfo {volume} {332}},\ \bibinfo {pages} {1173--1176}
  (\bibinfo {year} {2011})}\BibitemShut {NoStop}%
\bibitem [{\citenamefont {Depenbrock}\ \emph {et~al.}(2012)\citenamefont
  {Depenbrock}, \citenamefont {McCulloch},\ and\ \citenamefont
  {Schollw\"ock}}]{Shollwock}%
  \BibitemOpen
  \bibfield  {author} {\bibinfo {author} {\bibfnamefont {S.}~\bibnamefont
  {Depenbrock}}, \bibinfo {author} {\bibfnamefont {I.~P.}\ \bibnamefont
  {McCulloch}}, \ and\ \bibinfo {author} {\bibfnamefont {U.}~\bibnamefont
  {Schollw\"ock}},\ }\bibfield  {title} {\enquote {\bibinfo {title} {Nature of
  the spin-liquid ground state of the {$S=1/2$} {Heisenberg} model on the
  kagome lattice},}\ }\href {\doibase 10.1103/PhysRevLett.109.067201}
  {\bibfield  {journal} {\bibinfo  {journal} {Phys. Rev. Lett.}\ }\textbf
  {\bibinfo {volume} {109}},\ \bibinfo {pages} {067201} (\bibinfo {year}
  {2012})}\BibitemShut {NoStop}%
\bibitem [{\citenamefont {Han}\ \emph {et~al.}(2012)\citenamefont {Han},
  \citenamefont {Helton}, \citenamefont {Chu}, \citenamefont {Nocera},
  \citenamefont {Rodriguez-Rivera}, \citenamefont {Broholm},\ and\
  \citenamefont {Lee}}]{Han12}%
  \BibitemOpen
  \bibfield  {author} {\bibinfo {author} {\bibfnamefont {T.-H.}\ \bibnamefont
  {Han}}, \bibinfo {author} {\bibfnamefont {J.~S.}\ \bibnamefont {Helton}},
  \bibinfo {author} {\bibfnamefont {S.}~\bibnamefont {Chu}}, \bibinfo {author}
  {\bibfnamefont {D.~G.}\ \bibnamefont {Nocera}}, \bibinfo {author}
  {\bibfnamefont {J.~A.}\ \bibnamefont {Rodriguez-Rivera}}, \bibinfo {author}
  {\bibfnamefont {C.}~\bibnamefont {Broholm}}, \ and\ \bibinfo {author}
  {\bibfnamefont {Y.~S.}\ \bibnamefont {Lee}},\ }\bibfield  {title} {\enquote
  {\bibinfo {title} {Fractionalized excitations in the spin-liquid state of a
  kagome-lattice antiferromagnet},}\ }\href
  {http://dx.doi.org/10.1038/nature11659} {\bibfield  {journal} {\bibinfo
  {journal} {Nature (London)}\ }\textbf {\bibinfo {volume} {492}} (\bibinfo
  {year} {2012})}\BibitemShut {NoStop}%
\bibitem [{\citenamefont {Kitaev}(2006)}]{Kitaev2006}%
  \BibitemOpen
  \bibfield  {author} {\bibinfo {author} {\bibfnamefont {A.}~\bibnamefont
  {Kitaev}},\ }\bibfield  {title} {\enquote {\bibinfo {title} {{Anyons in an
  exactly solved model and beyond}},}\ }\href {\doibase
  http://dx.doi.org/10.1016/j.aop.2005.10.005} {\bibfield  {journal} {\bibinfo
  {journal} {Annals of Physics}\ }\textbf {\bibinfo {volume} {321}},\ \bibinfo
  {pages} {2 -- 111} (\bibinfo {year} {2006})}\BibitemShut {NoStop}%
\bibitem [{\citenamefont {Savary}\ and\ \citenamefont
  {Balents}(2017)}]{SavaryBalents}%
  \BibitemOpen
  \bibfield  {author} {\bibinfo {author} {\bibfnamefont {L.}~\bibnamefont
  {Savary}}\ and\ \bibinfo {author} {\bibfnamefont {L.}~\bibnamefont
  {Balents}},\ }\bibfield  {title} {\enquote {\bibinfo {title} {Quantum spin
  liquids: a review},}\ }\href {\doibase 10.1088/0034-4885/80/1/016502}
  {\bibfield  {journal} {\bibinfo  {journal} {Rep. Prog. Phys.}\ }\textbf
  {\bibinfo {volume} {80}},\ \bibinfo {pages} {016502} (\bibinfo {year}
  {2017})}\BibitemShut {NoStop}%
\bibitem [{\citenamefont {Gingras}\ and\ \citenamefont
  {McClarty}(2014)}]{gingras2014}%
  \BibitemOpen
  \bibfield  {author} {\bibinfo {author} {\bibfnamefont {M.~J.~P.}\
  \bibnamefont {Gingras}}\ and\ \bibinfo {author} {\bibfnamefont {P.~A.}\
  \bibnamefont {McClarty}},\ }\bibfield  {title} {\enquote {\bibinfo {title}
  {Quantum spin ice: a search for gapless quantum spin liquids in pyrochlore
  magnets},}\ }\href@noop {} {\bibfield  {journal} {\bibinfo  {journal} {Rep.
  Prog. Phys.}\ }\textbf {\bibinfo {volume} {77}},\ \bibinfo {pages} {056501}
  (\bibinfo {year} {2014})}\BibitemShut {NoStop}%
\bibitem [{\citenamefont {Castelnovo}\ \emph {et~al.}(2008)\citenamefont
  {Castelnovo}, \citenamefont {Moessner},\ and\ \citenamefont
  {Sondhi}}]{Castelnovo2008}%
  \BibitemOpen
  \bibfield  {author} {\bibinfo {author} {\bibfnamefont {C.}~\bibnamefont
  {Castelnovo}}, \bibinfo {author} {\bibfnamefont {R.}~\bibnamefont
  {Moessner}}, \ and\ \bibinfo {author} {\bibfnamefont {S.~L.}\ \bibnamefont
  {Sondhi}},\ }\bibfield  {title} {\enquote {\bibinfo {title} {Magnetic
  monopoles in spin ice},}\ }\href {\doibase
  http://dx.doi.org/10.1038/nature06433} {\bibfield  {journal} {\bibinfo
  {journal} {Nature}\ }\textbf {\bibinfo {volume} {451}},\ \bibinfo {pages}
  {42--45} (\bibinfo {year} {2008})}\BibitemShut {NoStop}%
\bibitem [{\citenamefont {Henley}(2010)}]{Henley2010}%
  \BibitemOpen
  \bibfield  {author} {\bibinfo {author} {\bibfnamefont {C.~L.}\ \bibnamefont
  {Henley}},\ }\bibfield  {title} {\enquote {\bibinfo {title} {The {Coulomb}
  phase in frustrated systems},}\ }\href {\doibase
  10.1146/annurev-conmatphys-070909-104138} {\bibfield  {journal} {\bibinfo
  {journal} {Ann. Rev. Condens. Matter Phys.}\ }\textbf {\bibinfo {volume}
  {1}},\ \bibinfo {pages} {179--210} (\bibinfo {year} {2010})}\BibitemShut
  {NoStop}%
\bibitem [{\citenamefont {Rehn}\ and\ \citenamefont
  {Moessner}(2016)}]{Rehn2016}%
  \BibitemOpen
  \bibfield  {author} {\bibinfo {author} {\bibfnamefont {J.}~\bibnamefont
  {Rehn}}\ and\ \bibinfo {author} {\bibfnamefont {R.}~\bibnamefont
  {Moessner}},\ }\bibfield  {title} {\enquote {\bibinfo {title} {Maxwell
  electromagnetism as an emergent phenomenon in condensed matter},}\ }\href
  {\doibase 10.1098/rsta.2016.0093} {\bibfield  {journal} {\bibinfo  {journal}
  {Phil. Trans. Royal Soc. London A}\ }\textbf {\bibinfo {volume} {374}}
  (\bibinfo {year} {2016}),\ 10.1098/rsta.2016.0093}\BibitemShut {NoStop}%
\bibitem [{\citenamefont {Cheong}\ and\ \citenamefont
  {Mostovoy}(2007)}]{mostovoy2007}%
  \BibitemOpen
  \bibfield  {author} {\bibinfo {author} {\bibfnamefont {S.-W.}\ \bibnamefont
  {Cheong}}\ and\ \bibinfo {author} {\bibfnamefont {M.}~\bibnamefont
  {Mostovoy}},\ }\bibfield  {title} {\enquote {\bibinfo {title} {Multiferroics:
  a magnetic twist for ferroelectricity},}\ }\href@noop {} {\bibfield
  {journal} {\bibinfo  {journal} {Nature Materials}\ }\textbf {\bibinfo
  {volume} {6}},\ \bibinfo {pages} {13--20} (\bibinfo {year}
  {2007})}\BibitemShut {NoStop}%
\bibitem [{\citenamefont {Arima}(2011)}]{arima2011}%
  \BibitemOpen
  \bibfield  {author} {\bibinfo {author} {\bibfnamefont {T.}~\bibnamefont
  {Arima}},\ }\bibfield  {title} {\enquote {\bibinfo {title} {Spin-driven
  ferroelectricity and magneto-electric effects in frustrated magnetic
  systems},}\ }\href@noop {} {\bibfield  {journal} {\bibinfo  {journal} {J.
  Phys. Soc. Jpn.}\ }\textbf {\bibinfo {volume} {80}},\ \bibinfo {pages}
  {052001} (\bibinfo {year} {2011})}\BibitemShut {NoStop}%
\bibitem [{\citenamefont {Tokura}\ \emph {et~al.}(2014)\citenamefont {Tokura},
  \citenamefont {Seki},\ and\ \citenamefont {Nagaosa}}]{tokura2014}%
  \BibitemOpen
  \bibfield  {author} {\bibinfo {author} {\bibfnamefont {Y.}~\bibnamefont
  {Tokura}}, \bibinfo {author} {\bibfnamefont {S.}~\bibnamefont {Seki}}, \ and\
  \bibinfo {author} {\bibfnamefont {N.}~\bibnamefont {Nagaosa}},\ }\bibfield
  {title} {\enquote {\bibinfo {title} {Multiferroics of spin origin},}\
  }\href@noop {} {\bibfield  {journal} {\bibinfo  {journal} {Rep. Prog. Phys.}\
  }\textbf {\bibinfo {volume} {77}},\ \bibinfo {pages} {076501} (\bibinfo
  {year} {2014})}\BibitemShut {NoStop}%
\bibitem [{Note1()}]{Note1}%
  \BibitemOpen
  \bibinfo {note} {This is mostly because of the well-known incompatibility of
  the five-fold symmetry with lattice periodicity.}\BibitemShut {Stop}%
\bibitem [{\citenamefont {Moessner}\ and\ \citenamefont
  {Sondhi}(2001)}]{Moessner2001}%
  \BibitemOpen
  \bibfield  {author} {\bibinfo {author} {\bibfnamefont {R.}~\bibnamefont
  {Moessner}}\ and\ \bibinfo {author} {\bibfnamefont {S.~L.}\ \bibnamefont
  {Sondhi}},\ }\bibfield  {title} {\enquote {\bibinfo {title} {Ising models of
  quantum frustration},}\ }\href {\doibase 10.1103/PhysRevB.63.224401}
  {\bibfield  {journal} {\bibinfo  {journal} {Phys. Rev. B}\ }\textbf {\bibinfo
  {volume} {63}},\ \bibinfo {pages} {224401} (\bibinfo {year}
  {2001})}\BibitemShut {NoStop}%
\bibitem [{\citenamefont {Urumov}(2002)}]{urumov2002}%
  \BibitemOpen
  \bibfield  {author} {\bibinfo {author} {\bibfnamefont {V.}~\bibnamefont
  {Urumov}},\ }\bibfield  {title} {\enquote {\bibinfo {title} {Exact solution
  of the {Ising} model on a pentagonal lattice},}\ }\href
  {http://stacks.iop.org/0305-4470/35/i=34/a=306} {\bibfield  {journal}
  {\bibinfo  {journal} {J. Phys. A}\ }\textbf {\bibinfo {volume} {35}},\
  \bibinfo {pages} {7317} (\bibinfo {year} {2002})}\BibitemShut {NoStop}%
\bibitem [{\citenamefont {Raman}\ \emph {et~al.}(2005)\citenamefont {Raman},
  \citenamefont {Moessner},\ and\ \citenamefont {Sondhi}}]{Raman2005}%
  \BibitemOpen
  \bibfield  {author} {\bibinfo {author} {\bibfnamefont {K.~S.}\ \bibnamefont
  {Raman}}, \bibinfo {author} {\bibfnamefont {R.}~\bibnamefont {Moessner}}, \
  and\ \bibinfo {author} {\bibfnamefont {S.~L.}\ \bibnamefont {Sondhi}},\
  }\bibfield  {title} {\enquote {\bibinfo {title} {{SU(2)}-invariant
  spin-$\frac{1}{2}$ {Hamiltonians} with resonating and other valence bond
  phases},}\ }\href {\doibase 10.1103/PhysRevB.72.064413} {\bibfield  {journal}
  {\bibinfo  {journal} {Phys. Rev. B}\ }\textbf {\bibinfo {volume} {72}},\
  \bibinfo {pages} {064413} (\bibinfo {year} {2005})}\BibitemShut {NoStop}%
\bibitem [{\citenamefont {Jagannathan}\ \emph {et~al.}(2011)\citenamefont
  {Jagannathan}, \citenamefont {Motz},\ and\ \citenamefont
  {Vedmedenko}}]{Jagannathan2011}%
  \BibitemOpen
  \bibfield  {author} {\bibinfo {author} {\bibfnamefont {A.}~\bibnamefont
  {Jagannathan}}, \bibinfo {author} {\bibfnamefont {B.}~\bibnamefont {Motz}}, \
  and\ \bibinfo {author} {\bibfnamefont {E.}~\bibnamefont {Vedmedenko}},\
  }\bibfield  {title} {\enquote {\bibinfo {title} {Novel properties of
  frustrated low-dimensional magnets with pentagonal symmetry},}\ }\href
  {\doibase 10.1080/14786435.2010.539578} {\bibfield  {journal} {\bibinfo
  {journal} {Phil. Mag.}\ }\textbf {\bibinfo {volume} {91}},\ \bibinfo {pages}
  {2765--2772} (\bibinfo {year} {2011})}\BibitemShut {NoStop}%
\bibitem [{\citenamefont {Ralko}(2011)}]{ralko2011}%
  \BibitemOpen
  \bibfield  {author} {\bibinfo {author} {\bibfnamefont {A.}~\bibnamefont
  {Ralko}},\ }\bibfield  {title} {\enquote {\bibinfo {title} {Phase diagram of
  the {Cairo} pentagonal {$XXZ$} spin-$\frac{1}{2}$ magnet under a magnetic
  field},}\ }\href {\doibase 10.1103/PhysRevB.84.184434} {\bibfield  {journal}
  {\bibinfo  {journal} {Phys. Rev. B}\ }\textbf {\bibinfo {volume} {84}},\
  \bibinfo {pages} {184434} (\bibinfo {year} {2011})}\BibitemShut {NoStop}%
\bibitem [{\citenamefont {Rojas}\ \emph {et~al.}(2012)\citenamefont {Rojas},
  \citenamefont {Rojas},\ and\ \citenamefont {de~Souza}}]{rojas2012}%
  \BibitemOpen
  \bibfield  {author} {\bibinfo {author} {\bibfnamefont {M.}~\bibnamefont
  {Rojas}}, \bibinfo {author} {\bibfnamefont {O.}~\bibnamefont {Rojas}}, \ and\
  \bibinfo {author} {\bibfnamefont {S.~M.}\ \bibnamefont {de~Souza}},\
  }\bibfield  {title} {\enquote {\bibinfo {title} {Frustrated {Ising} model on
  the {Cairo} pentagonal lattice},}\ }\href {\doibase
  10.1103/PhysRevE.86.051116} {\bibfield  {journal} {\bibinfo  {journal} {Phys.
  Rev. E}\ }\textbf {\bibinfo {volume} {86}},\ \bibinfo {pages} {051116}
  (\bibinfo {year} {2012})}\BibitemShut {NoStop}%
\bibitem [{\citenamefont {Rousochatzakis}\ \emph {et~al.}(2012)\citenamefont
  {Rousochatzakis}, \citenamefont {L\"auchli},\ and\ \citenamefont
  {Moessner}}]{rousochatzakis2012}%
  \BibitemOpen
  \bibfield  {author} {\bibinfo {author} {\bibfnamefont {I.}~\bibnamefont
  {Rousochatzakis}}, \bibinfo {author} {\bibfnamefont {A.~M.}\ \bibnamefont
  {L\"auchli}}, \ and\ \bibinfo {author} {\bibfnamefont {R.}~\bibnamefont
  {Moessner}},\ }\bibfield  {title} {\enquote {\bibinfo {title} {Quantum
  magnetism on the {Cairo} pentagonal lattice},}\ }\href {\doibase
  10.1103/PhysRevB.85.104415} {\bibfield  {journal} {\bibinfo  {journal} {Phys.
  Rev. B}\ }\textbf {\bibinfo {volume} {85}},\ \bibinfo {pages} {104415}
  (\bibinfo {year} {2012})}\BibitemShut {NoStop}%
\bibitem [{\citenamefont {Pchelkina}\ and\ \citenamefont
  {Streltsov}(2013)}]{pchelkina2013}%
  \BibitemOpen
  \bibfield  {author} {\bibinfo {author} {\bibfnamefont {Z.~V.}\ \bibnamefont
  {Pchelkina}}\ and\ \bibinfo {author} {\bibfnamefont {S.~V.}\ \bibnamefont
  {Streltsov}},\ }\bibfield  {title} {\enquote {\bibinfo {title} {\textit{Ab
  initio} investigation of the exchange interactions in
  {Bi${}_{2}$Fe${}_{4}$O${}_{9}$}: The {Cairo} pentagonal lattice compound},}\
  }\href {\doibase 10.1103/PhysRevB.88.054424} {\bibfield  {journal} {\bibinfo
  {journal} {Phys. Rev. B}\ }\textbf {\bibinfo {volume} {88}},\ \bibinfo
  {pages} {054424} (\bibinfo {year} {2013})}\BibitemShut {NoStop}%
\bibitem [{\citenamefont {Nakano}\ \emph {et~al.}(2014)\citenamefont {Nakano},
  \citenamefont {Isoda},\ and\ \citenamefont {Sakai}}]{Nakano2014}%
  \BibitemOpen
  \bibfield  {author} {\bibinfo {author} {\bibfnamefont {H.}~\bibnamefont
  {Nakano}}, \bibinfo {author} {\bibfnamefont {M.}~\bibnamefont {Isoda}}, \
  and\ \bibinfo {author} {\bibfnamefont {T.}~\bibnamefont {Sakai}},\ }\bibfield
   {title} {\enquote {\bibinfo {title} {Magnetization process of the {$S=1/2$}
  {Heisenberg} antiferromagnet on the {Cairo} pentagon lattice},}\ }\href
  {\doibase 10.7566/JPSJ.83.053702} {\bibfield  {journal} {\bibinfo  {journal}
  {J. Phys. Soc. Jpn.}\ }\textbf {\bibinfo {volume} {83}},\ \bibinfo {pages}
  {053702} (\bibinfo {year} {2014})}\BibitemShut {NoStop}%
\bibitem [{\citenamefont {Isoda}\ \emph {et~al.}(2014)\citenamefont {Isoda},
  \citenamefont {Nakano},\ and\ \citenamefont {Sakai}}]{Isoda2014}%
  \BibitemOpen
  \bibfield  {author} {\bibinfo {author} {\bibfnamefont {M.}~\bibnamefont
  {Isoda}}, \bibinfo {author} {\bibfnamefont {H.}~\bibnamefont {Nakano}}, \
  and\ \bibinfo {author} {\bibfnamefont {T.}~\bibnamefont {Sakai}},\ }\bibfield
   {title} {\enquote {\bibinfo {title} {Frustration-induced magnetic properties
  of the spin-1/2 {Heisenberg} antiferromagnet on the {Cairo} pentagon
  lattice},}\ }\href {\doibase 10.7566/JPSJ.83.084710} {\bibfield  {journal}
  {\bibinfo  {journal} {J. Phys. Soc. Jpn.}\ }\textbf {\bibinfo {volume}
  {83}},\ \bibinfo {pages} {084710} (\bibinfo {year} {2014})}\BibitemShut
  {NoStop}%
\bibitem [{\citenamefont {Shamir}\ \emph {et~al.}(1978)\citenamefont {Shamir},
  \citenamefont {Gurewitz},\ and\ \citenamefont {Shaked}}]{shamir1978}%
  \BibitemOpen
  \bibfield  {author} {\bibinfo {author} {\bibfnamefont {N.}~\bibnamefont
  {Shamir}}, \bibinfo {author} {\bibfnamefont {E.}~\bibnamefont {Gurewitz}}, \
  and\ \bibinfo {author} {\bibfnamefont {H.}~\bibnamefont {Shaked}},\
  }\bibfield  {title} {\enquote {\bibinfo {title} {{The magnetic structure of
  {Bi$_2$Fe$_4$O$_9$} {--} analysis of neutron diffraction measurements}},}\
  }\href {\doibase 10.1107/S0567739478001412} {\bibfield  {journal} {\bibinfo
  {journal} {Acta Cryst.}\ }\textbf {\bibinfo {volume} {A34}},\ \bibinfo
  {pages} {662--666} (\bibinfo {year} {1978})}\BibitemShut {NoStop}%
\bibitem [{\citenamefont {Singh}\ \emph {et~al.}(2008)\citenamefont {Singh},
  \citenamefont {Kaushik}, \citenamefont {Kumar}, \citenamefont {Mishra},
  \citenamefont {Venimadhav}, \citenamefont {Siruguri},\ and\ \citenamefont
  {Patnaik}}]{Singh2008}%
  \BibitemOpen
  \bibfield  {author} {\bibinfo {author} {\bibfnamefont {A.~K.}\ \bibnamefont
  {Singh}}, \bibinfo {author} {\bibfnamefont {S.~D.}\ \bibnamefont {Kaushik}},
  \bibinfo {author} {\bibfnamefont {B.}~\bibnamefont {Kumar}}, \bibinfo
  {author} {\bibfnamefont {P.~K.}\ \bibnamefont {Mishra}}, \bibinfo {author}
  {\bibfnamefont {A.}~\bibnamefont {Venimadhav}}, \bibinfo {author}
  {\bibfnamefont {V.}~\bibnamefont {Siruguri}}, \ and\ \bibinfo {author}
  {\bibfnamefont {S.}~\bibnamefont {Patnaik}},\ }\bibfield  {title} {\enquote
  {\bibinfo {title} {Substantial magnetoelectric coupling near room temperature
  in {Bi$_2$Fe$_4$O$_9$}},}\ }\href {http://dx.doi.org/10.1063/1.2905815}
  {\bibfield  {journal} {\bibinfo  {journal} {Appl. Phys. Lett.}\ }\textbf
  {\bibinfo {volume} {92}},\ \bibinfo {pages} {132910} (\bibinfo {year}
  {2008})}\BibitemShut {NoStop}%
\bibitem [{\citenamefont {Ressouche}\ \emph {et~al.}(2009)\citenamefont
  {Ressouche}, \citenamefont {Simonet}, \citenamefont {Canals}, \citenamefont
  {Gospodinov},\ and\ \citenamefont {Skumryev}}]{ressouche2009}%
  \BibitemOpen
  \bibfield  {author} {\bibinfo {author} {\bibfnamefont {E.}~\bibnamefont
  {Ressouche}}, \bibinfo {author} {\bibfnamefont {V.}~\bibnamefont {Simonet}},
  \bibinfo {author} {\bibfnamefont {B.}~\bibnamefont {Canals}}, \bibinfo
  {author} {\bibfnamefont {M.}~\bibnamefont {Gospodinov}}, \ and\ \bibinfo
  {author} {\bibfnamefont {V.}~\bibnamefont {Skumryev}},\ }\bibfield  {title}
  {\enquote {\bibinfo {title} {Magnetic frustration in an iron-based {Cairo}
  pentagonal lattice},}\ }\href {\doibase 10.1103/PhysRevLett.103.267204}
  {\bibfield  {journal} {\bibinfo  {journal} {Phys. Rev. Lett.}\ }\textbf
  {\bibinfo {volume} {103}},\ \bibinfo {pages} {267204} (\bibinfo {year}
  {2009})}\BibitemShut {NoStop}%
\bibitem [{\citenamefont {Iliev}\ \emph {et~al.}(2010)\citenamefont {Iliev},
  \citenamefont {Litvinchuk}, \citenamefont {Hadjiev}, \citenamefont
  {Gospodinov}, \citenamefont {Skumryev},\ and\ \citenamefont
  {Ressouche}}]{iliev2010}%
  \BibitemOpen
  \bibfield  {author} {\bibinfo {author} {\bibfnamefont {M.~N.}\ \bibnamefont
  {Iliev}}, \bibinfo {author} {\bibfnamefont {A.~P.}\ \bibnamefont
  {Litvinchuk}}, \bibinfo {author} {\bibfnamefont {V.~G.}\ \bibnamefont
  {Hadjiev}}, \bibinfo {author} {\bibfnamefont {M.~M.}\ \bibnamefont
  {Gospodinov}}, \bibinfo {author} {\bibfnamefont {V.}~\bibnamefont
  {Skumryev}}, \ and\ \bibinfo {author} {\bibfnamefont {E.}~\bibnamefont
  {Ressouche}},\ }\bibfield  {title} {\enquote {\bibinfo {title} {Phonon and
  magnon scattering of antiferromagnetic {Bi$_2$Fe$_4$O$_9$}},}\ }\href
  {\doibase 10.1103/PhysRevB.81.024302} {\bibfield  {journal} {\bibinfo
  {journal} {Phys. Rev. B}\ }\textbf {\bibinfo {volume} {81}},\ \bibinfo
  {pages} {024302} (\bibinfo {year} {2010})}\BibitemShut {NoStop}%
\bibitem [{\citenamefont {Abakumov}\ \emph {et~al.}(2013)\citenamefont
  {Abakumov}, \citenamefont {Batuk}, \citenamefont {Tsirlin}, \citenamefont
  {Prescher}, \citenamefont {Dubrovinsky}, \citenamefont {Sheptyakov},
  \citenamefont {Schnelle}, \citenamefont {Hadermann},\ and\ \citenamefont
  {{Van Tendeloo}}}]{abakumov2013}%
  \BibitemOpen
  \bibfield  {author} {\bibinfo {author} {\bibfnamefont {A.~M.}\ \bibnamefont
  {Abakumov}}, \bibinfo {author} {\bibfnamefont {D.}~\bibnamefont {Batuk}},
  \bibinfo {author} {\bibfnamefont {A.~A.}\ \bibnamefont {Tsirlin}}, \bibinfo
  {author} {\bibfnamefont {C.}~\bibnamefont {Prescher}}, \bibinfo {author}
  {\bibfnamefont {L.}~\bibnamefont {Dubrovinsky}}, \bibinfo {author}
  {\bibfnamefont {D.~V.}\ \bibnamefont {Sheptyakov}}, \bibinfo {author}
  {\bibfnamefont {W.}~\bibnamefont {Schnelle}}, \bibinfo {author}
  {\bibfnamefont {J.}~\bibnamefont {Hadermann}}, \ and\ \bibinfo {author}
  {\bibfnamefont {G.}~\bibnamefont {{Van Tendeloo}}},\ }\bibfield  {title}
  {\enquote {\bibinfo {title} {Frustrated pentagonal {Cairo} lattice in the
  non-collinear antiferromagnet {Bi$_4$Fe$_5$O$_{13}$F}},}\ }\href {\doibase
  10.1103/PhysRevB.87.024423} {\bibfield  {journal} {\bibinfo  {journal} {Phys.
  Rev. B}\ }\textbf {\bibinfo {volume} {87}},\ \bibinfo {pages} {024423}
  (\bibinfo {year} {2013})}\BibitemShut {NoStop}%
\bibitem [{\citenamefont {Rozova}\ \emph {et~al.}(2016)\citenamefont {Rozova},
  \citenamefont {Grigoriev}, \citenamefont {Bobrikov}, \citenamefont
  {Filimonov}, \citenamefont {Zakharov}, \citenamefont {Volkova}, \citenamefont
  {Vasiliev}, \citenamefont {Antipov}, \citenamefont {Tsirlin},\ and\
  \citenamefont {Abakumov}}]{Rozova2016}%
  \BibitemOpen
  \bibfield  {author} {\bibinfo {author} {\bibfnamefont {M.~G.}\ \bibnamefont
  {Rozova}}, \bibinfo {author} {\bibfnamefont {V.~V.}\ \bibnamefont
  {Grigoriev}}, \bibinfo {author} {\bibfnamefont {I.~A.}\ \bibnamefont
  {Bobrikov}}, \bibinfo {author} {\bibfnamefont {D.~S.}\ \bibnamefont
  {Filimonov}}, \bibinfo {author} {\bibfnamefont {K.~V.}\ \bibnamefont
  {Zakharov}}, \bibinfo {author} {\bibfnamefont {O.~S.}\ \bibnamefont
  {Volkova}}, \bibinfo {author} {\bibfnamefont {A.~N.}\ \bibnamefont
  {Vasiliev}}, \bibinfo {author} {\bibfnamefont {E.~V.}\ \bibnamefont
  {Antipov}}, \bibinfo {author} {\bibfnamefont {A.~A.}\ \bibnamefont
  {Tsirlin}}, \ and\ \bibinfo {author} {\bibfnamefont {A.~M.}\ \bibnamefont
  {Abakumov}},\ }\bibfield  {title} {\enquote {\bibinfo {title} {Synthesis{,}
  structure and magnetic ordering of the mullite-type
  {Bi$_2$Fe$_{4-x}$Cr$_x$O$_9$} solid solutions with a frustrated pentagonal
  {Cairo} lattice},}\ }\href {\doibase 10.1039/C5DT04296H} {\bibfield
  {journal} {\bibinfo  {journal} {Dalton Trans.}\ }\textbf {\bibinfo {volume}
  {45}},\ \bibinfo {pages} {1192--1200} (\bibinfo {year} {2016})}\BibitemShut
  {NoStop}%
\bibitem [{\citenamefont {Mohapatra}\ \emph {et~al.}(2016)\citenamefont
  {Mohapatra}, \citenamefont {Swain}, \citenamefont {Yadav}, \citenamefont
  {Kaushik},\ and\ \citenamefont {Singh}}]{Mohapatra2016}%
  \BibitemOpen
  \bibfield  {author} {\bibinfo {author} {\bibfnamefont {S.~R.}\ \bibnamefont
  {Mohapatra}}, \bibinfo {author} {\bibfnamefont {A.}~\bibnamefont {Swain}},
  \bibinfo {author} {\bibfnamefont {C.~S.}\ \bibnamefont {Yadav}}, \bibinfo
  {author} {\bibfnamefont {S.~D.}\ \bibnamefont {Kaushik}}, \ and\ \bibinfo
  {author} {\bibfnamefont {A.~K.}\ \bibnamefont {Singh}},\ }\bibfield  {title}
  {\enquote {\bibinfo {title} {Unequivocal evidence of enhanced
  magnetodielectric coupling in {Gd$^{3+}$} substituted multiferroic
  {Bi$_2$Fe$_4$O$_9$}},}\ }\href {\doibase 10.1039/C6RA24525K} {\bibfield
  {journal} {\bibinfo  {journal} {RSC Adv.}\ }\textbf {\bibinfo {volume} {6}},\
  \bibinfo {pages} {112282--112291} (\bibinfo {year} {2016})}\BibitemShut
  {NoStop}%
\bibitem [{\citenamefont {Saito}\ and\ \citenamefont {Kohn}(1995)}]{Saito1995}%
  \BibitemOpen
  \bibfield  {author} {\bibinfo {author} {\bibfnamefont {K.}~\bibnamefont
  {Saito}}\ and\ \bibinfo {author} {\bibfnamefont {K.}~\bibnamefont {Kohn}},\
  }\bibfield  {title} {\enquote {\bibinfo {title} {Magnetoelectric effect and
  low-temperature phase transitions of {TbMn$_2$O$_5$}},}\ }\href
  {http://stacks.iop.org/0953-8984/7/i=14/a=022} {\bibfield  {journal}
  {\bibinfo  {journal} {J. Phys.: Condens. Matter}\ }\textbf {\bibinfo {volume}
  {7}},\ \bibinfo {pages} {2855} (\bibinfo {year} {1995})}\BibitemShut
  {NoStop}%
\bibitem [{\citenamefont {Hur}\ \emph {et~al.}(2004)\citenamefont {Hur},
  \citenamefont {Park}, \citenamefont {Sharma}, \citenamefont {Ahn},
  \citenamefont {Guha},\ and\ \citenamefont {Cheong}}]{Hur2004}%
  \BibitemOpen
  \bibfield  {author} {\bibinfo {author} {\bibfnamefont {N.}~\bibnamefont
  {Hur}}, \bibinfo {author} {\bibfnamefont {S.}~\bibnamefont {Park}}, \bibinfo
  {author} {\bibfnamefont {P.~A.}\ \bibnamefont {Sharma}}, \bibinfo {author}
  {\bibfnamefont {J.~S.}\ \bibnamefont {Ahn}}, \bibinfo {author} {\bibfnamefont
  {S.}~\bibnamefont {Guha}}, \ and\ \bibinfo {author} {\bibfnamefont {S-W.}\
  \bibnamefont {Cheong}},\ }\bibfield  {title} {\enquote {\bibinfo {title}
  {Electric polarization reversal and memory in a multiferroic material induced
  by magnetic fields},}\ }\href {\doibase doi:10.1038/nature02572} {\bibfield
  {journal} {\bibinfo  {journal} {Nature}\ }\textbf {\bibinfo {volume} {429}},\
  \bibinfo {pages} {392} (\bibinfo {year} {2004})}\BibitemShut {NoStop}%
\bibitem [{\citenamefont {Blake}\ \emph {et~al.}(2005)\citenamefont {Blake},
  \citenamefont {Chapon}, \citenamefont {Radaelli}, \citenamefont {Park},
  \citenamefont {Hur}, \citenamefont {Cheong},\ and\ \citenamefont
  {Rodr\'{\i}guez-Carvajal}}]{blake2005}%
  \BibitemOpen
  \bibfield  {author} {\bibinfo {author} {\bibfnamefont {G.~R.}\ \bibnamefont
  {Blake}}, \bibinfo {author} {\bibfnamefont {L.~C.}\ \bibnamefont {Chapon}},
  \bibinfo {author} {\bibfnamefont {P.~G.}\ \bibnamefont {Radaelli}}, \bibinfo
  {author} {\bibfnamefont {S.}~\bibnamefont {Park}}, \bibinfo {author}
  {\bibfnamefont {N.}~\bibnamefont {Hur}}, \bibinfo {author} {\bibfnamefont
  {S-W.}\ \bibnamefont {Cheong}}, \ and\ \bibinfo {author} {\bibfnamefont
  {J.}~\bibnamefont {Rodr\'{\i}guez-Carvajal}},\ }\bibfield  {title} {\enquote
  {\bibinfo {title} {Spin structure and magnetic frustration in multiferroic
  {$R$Mn$_2$O$_5$} ({$R$ = Tb, Ho, Dy})},}\ }\href {\doibase
  10.1103/PhysRevB.71.214402} {\bibfield  {journal} {\bibinfo  {journal} {Phys.
  Rev. B}\ }\textbf {\bibinfo {volume} {71}},\ \bibinfo {pages} {214402}
  (\bibinfo {year} {2005})}\BibitemShut {NoStop}%
\bibitem [{\citenamefont {Vecchini}\ \emph {et~al.}(2008)\citenamefont
  {Vecchini}, \citenamefont {Chapon}, \citenamefont {Brown}, \citenamefont
  {Chatterji}, \citenamefont {Park}, \citenamefont {Cheong},\ and\
  \citenamefont {Radaelli}}]{vecchini2008}%
  \BibitemOpen
  \bibfield  {author} {\bibinfo {author} {\bibfnamefont {C.}~\bibnamefont
  {Vecchini}}, \bibinfo {author} {\bibfnamefont {L.~C.}\ \bibnamefont
  {Chapon}}, \bibinfo {author} {\bibfnamefont {P.~J.}\ \bibnamefont {Brown}},
  \bibinfo {author} {\bibfnamefont {T.}~\bibnamefont {Chatterji}}, \bibinfo
  {author} {\bibfnamefont {S.}~\bibnamefont {Park}}, \bibinfo {author}
  {\bibfnamefont {S-W.}\ \bibnamefont {Cheong}}, \ and\ \bibinfo {author}
  {\bibfnamefont {P.~G.}\ \bibnamefont {Radaelli}},\ }\bibfield  {title}
  {\enquote {\bibinfo {title} {Commensurate magnetic structures of
  {$R$Mn$_2$O$_{5}$} ({$R$ = Y, Ho, Bi}) determined by single-crystal neutron
  diffraction},}\ }\href {\doibase 10.1103/PhysRevB.77.134434} {\bibfield
  {journal} {\bibinfo  {journal} {Phys. Rev. B}\ }\textbf {\bibinfo {volume}
  {77}},\ \bibinfo {pages} {134434} (\bibinfo {year} {2008})}\BibitemShut
  {NoStop}%
\bibitem [{\citenamefont {Harris}\ \emph {et~al.}(2008)\citenamefont {Harris},
  \citenamefont {Aharony},\ and\ \citenamefont {Entin-Wohlman}}]{harris2008}%
  \BibitemOpen
  \bibfield  {author} {\bibinfo {author} {\bibfnamefont {A.~B.}\ \bibnamefont
  {Harris}}, \bibinfo {author} {\bibfnamefont {A.}~\bibnamefont {Aharony}}, \
  and\ \bibinfo {author} {\bibfnamefont {O.}~\bibnamefont {Entin-Wohlman}},\
  }\bibfield  {title} {\enquote {\bibinfo {title} {Order parameters and phase
  diagram of multiferroic {$R{\mathrm{Mn}}_{2}{\mathrm{O}}_{5}$}},}\ }\href
  {\doibase 10.1103/PhysRevLett.100.217202} {\bibfield  {journal} {\bibinfo
  {journal} {Phys. Rev. Lett.}\ }\textbf {\bibinfo {volume} {100}},\ \bibinfo
  {pages} {217202} (\bibinfo {year} {2008})}\BibitemShut {NoStop}%
\bibitem [{\citenamefont {Sushkov}\ \emph {et~al.}(2008)\citenamefont
  {Sushkov}, \citenamefont {Mostovoy}, \citenamefont {{Vald\'es Aguilar}},
  \citenamefont {Cheong},\ and\ \citenamefont {Drew}}]{Sushkov2008}%
  \BibitemOpen
  \bibfield  {author} {\bibinfo {author} {\bibfnamefont {A.~B.}\ \bibnamefont
  {Sushkov}}, \bibinfo {author} {\bibfnamefont {M.}~\bibnamefont {Mostovoy}},
  \bibinfo {author} {\bibfnamefont {R.}~\bibnamefont {{Vald\'es Aguilar}}},
  \bibinfo {author} {\bibfnamefont {S.-W.}\ \bibnamefont {Cheong}}, \ and\
  \bibinfo {author} {\bibfnamefont {H.~D.}\ \bibnamefont {Drew}},\ }\bibfield
  {title} {\enquote {\bibinfo {title} {Electromagnons in multiferroic
  {RMn$_2$O$_5$} compounds and their microscopic origin},}\ }\href
  {http://stacks.iop.org/0953-8984/20/i=43/a=434210} {\bibfield  {journal}
  {\bibinfo  {journal} {J. Phys.: Condens. Matter}\ }\textbf {\bibinfo {volume}
  {20}},\ \bibinfo {pages} {434210} (\bibinfo {year} {2008})}\BibitemShut
  {NoStop}%
\bibitem [{\citenamefont {Kim}\ \emph {et~al.}(2009)\citenamefont {Kim},
  \citenamefont {Haam}, \citenamefont {Oh}, \citenamefont {Park}, \citenamefont
  {Cheong}, \citenamefont {Sharma}, \citenamefont {Jaime}, \citenamefont
  {Harrison}, \citenamefont {Han}, \citenamefont {Jeon}, \citenamefont
  {Coleman},\ and\ \citenamefont {Kim}}]{Kim2009}%
  \BibitemOpen
  \bibfield  {author} {\bibinfo {author} {\bibfnamefont {J.~W.}\ \bibnamefont
  {Kim}}, \bibinfo {author} {\bibfnamefont {S.~Y.}\ \bibnamefont {Haam}},
  \bibinfo {author} {\bibfnamefont {Y.~S.}\ \bibnamefont {Oh}}, \bibinfo
  {author} {\bibfnamefont {S.}~\bibnamefont {Park}}, \bibinfo {author}
  {\bibfnamefont {S.-W.}\ \bibnamefont {Cheong}}, \bibinfo {author}
  {\bibfnamefont {P.~A.}\ \bibnamefont {Sharma}}, \bibinfo {author}
  {\bibfnamefont {M.}~\bibnamefont {Jaime}}, \bibinfo {author} {\bibfnamefont
  {N.}~\bibnamefont {Harrison}}, \bibinfo {author} {\bibfnamefont {Jung~Hoon}\
  \bibnamefont {Han}}, \bibinfo {author} {\bibfnamefont {G.-S.}\ \bibnamefont
  {Jeon}}, \bibinfo {author} {\bibfnamefont {P.}~\bibnamefont {Coleman}}, \
  and\ \bibinfo {author} {\bibfnamefont {K.~H.}\ \bibnamefont {Kim}},\
  }\bibfield  {title} {\enquote {\bibinfo {title} {Observation of a
  multiferroic critical end point},}\ }\href {\doibase 10.1073/pnas.0907589106}
  {\bibfield  {journal} {\bibinfo  {journal} {PNAS}\ }\textbf {\bibinfo
  {volume} {106}},\ \bibinfo {pages} {15573--15576} (\bibinfo {year}
  {2009})}\BibitemShut {NoStop}%
\bibitem [{\citenamefont {Cao}\ \emph {et~al.}(2009)\citenamefont {Cao},
  \citenamefont {Guo}, \citenamefont {Vanderbilt},\ and\ \citenamefont
  {He}}]{cao2009}%
  \BibitemOpen
  \bibfield  {author} {\bibinfo {author} {\bibfnamefont {K.}~\bibnamefont
  {Cao}}, \bibinfo {author} {\bibfnamefont {G.-C.}\ \bibnamefont {Guo}},
  \bibinfo {author} {\bibfnamefont {D.}~\bibnamefont {Vanderbilt}}, \ and\
  \bibinfo {author} {\bibfnamefont {L.}~\bibnamefont {He}},\ }\bibfield
  {title} {\enquote {\bibinfo {title} {First-principles modeling of
  multiferroic {RMn$_2$O$_5$}},}\ }\href {\doibase
  10.1103/PhysRevLett.103.257201} {\bibfield  {journal} {\bibinfo  {journal}
  {Phys. Rev. Lett.}\ }\textbf {\bibinfo {volume} {103}},\ \bibinfo {pages}
  {257201} (\bibinfo {year} {2009})}\BibitemShut {NoStop}%
\bibitem [{\citenamefont {Momma}\ and\ \citenamefont {Izumi}(2011)}]{vesta}%
  \BibitemOpen
  \bibfield  {author} {\bibinfo {author} {\bibfnamefont {K.}~\bibnamefont
  {Momma}}\ and\ \bibinfo {author} {\bibfnamefont {F.}~\bibnamefont {Izumi}},\
  }\bibfield  {title} {\enquote {\bibinfo {title} {\textit{VESTA} 3 for
  three-dimensional visualization of crystal, volumetric and morphology
  data},}\ }\href@noop {} {\bibfield  {journal} {\bibinfo  {journal} {J. Appl.
  Crystallogr.}\ }\textbf {\bibinfo {volume} {44}},\ \bibinfo {pages}
  {1272--1276} (\bibinfo {year} {2011})}\BibitemShut {NoStop}%
\bibitem [{\citenamefont {Nath}\ \emph {et~al.}(2014)\citenamefont {Nath},
  \citenamefont {Ranjith}, \citenamefont {Sichelschmidt}, \citenamefont
  {Baenitz}, \citenamefont {Skourski}, \citenamefont {Alet}, \citenamefont
  {Rousochatzakis},\ and\ \citenamefont {Tsirlin}}]{CuP2O6}%
  \BibitemOpen
  \bibfield  {author} {\bibinfo {author} {\bibfnamefont {R.}~\bibnamefont
  {Nath}}, \bibinfo {author} {\bibfnamefont {K.~M.}\ \bibnamefont {Ranjith}},
  \bibinfo {author} {\bibfnamefont {J.}~\bibnamefont {Sichelschmidt}}, \bibinfo
  {author} {\bibfnamefont {M.}~\bibnamefont {Baenitz}}, \bibinfo {author}
  {\bibfnamefont {Y.}~\bibnamefont {Skourski}}, \bibinfo {author}
  {\bibfnamefont {F.}~\bibnamefont {Alet}}, \bibinfo {author} {\bibfnamefont
  {I.}~\bibnamefont {Rousochatzakis}}, \ and\ \bibinfo {author} {\bibfnamefont
  {A.~A.}\ \bibnamefont {Tsirlin}},\ }\bibfield  {title} {\enquote {\bibinfo
  {title} {Hindered magnetic order from mixed dimensionalities in
  {${\text{CuP}}_{2}{\text{O}}_{6}$}},}\ }\href {\doibase
  10.1103/PhysRevB.89.014407} {\bibfield  {journal} {\bibinfo  {journal} {Phys.
  Rev. B}\ }\textbf {\bibinfo {volume} {89}},\ \bibinfo {pages} {014407}
  (\bibinfo {year} {2014})}\BibitemShut {NoStop}%
\bibitem [{\citenamefont {Pet{\u r}{\'\i}{\u c}ek}\ \emph
  {et~al.}(2014)\citenamefont {Pet{\u r}{\'\i}{\u c}ek}, \citenamefont {Du{\u
  s}ek},\ and\ \citenamefont {Palatinus}}]{jana2006}%
  \BibitemOpen
  \bibfield  {author} {\bibinfo {author} {\bibfnamefont {V.}~\bibnamefont
  {Pet{\u r}{\'\i}{\u c}ek}}, \bibinfo {author} {\bibfnamefont
  {M.}~\bibnamefont {Du{\u s}ek}}, \ and\ \bibinfo {author} {\bibfnamefont
  {L.}~\bibnamefont {Palatinus}},\ }\bibfield  {title} {\enquote {\bibinfo
  {title} {Crystallographic computing system {JANA2006}: General features},}\
  }\href {\doibase 10.1515/zkri-2014-1737} {\bibfield  {journal} {\bibinfo
  {journal} {Z. Krist.}\ }\textbf {\bibinfo {volume} {229}},\ \bibinfo {pages}
  {345--352} (\bibinfo {year} {2014})}\BibitemShut {NoStop}%
\bibitem [{\citenamefont {Campbell}\ \emph {et~al.}(2006)\citenamefont
  {Campbell}, \citenamefont {Stokes}, \citenamefont {Tanner},\ and\
  \citenamefont {Hatch}}]{isodistort}%
  \BibitemOpen
  \bibfield  {author} {\bibinfo {author} {\bibfnamefont {B.~J.}\ \bibnamefont
  {Campbell}}, \bibinfo {author} {\bibfnamefont {H.~T.}\ \bibnamefont
  {Stokes}}, \bibinfo {author} {\bibfnamefont {D.~E.}\ \bibnamefont {Tanner}},
  \ and\ \bibinfo {author} {\bibfnamefont {D.~M.}\ \bibnamefont {Hatch}},\
  }\bibfield  {title} {\enquote {\bibinfo {title} {{{\it ISODISPLACE}: a
  web-based tool for exploring structural distortions}},}\ }\href {\doibase
  10.1107/S0021889806014075} {\bibfield  {journal} {\bibinfo  {journal} {J.
  Appl. Cryst.}\ }\textbf {\bibinfo {volume} {39}},\ \bibinfo {pages}
  {607--614} (\bibinfo {year} {2006})}\BibitemShut {NoStop}%
\bibitem [{Note2()}]{Note2}%
  \BibitemOpen
  \bibinfo {note} {We have also tested a different magnetic structure solution
  that would impose largely non-collinear spins in the spirit of phase I. To
  this end, the space group $P_Anna$ was used ($P_Ccnn$ maintaining the crystal
  axes of $P_C4_2/n$). This solution, however, demonstrated an inferior fit of
  the magnetic reflections ($R_{\protect \rm mag}=0.030$ at $T=65$\protect
  \tmspace +\thinmuskip {.1667em}K) because spin directions for the Fe3 atoms
  were constrained by symmetry. A better fit might be achievable with the
  $P_c2/c$ magnetic space group, but it contains 10 independent
  crystallographic positions for the Fe atoms that renders the solution
  intractable.}\BibitemShut {Stop}%
\bibitem [{\citenamefont {Perdew}\ \emph {et~al.}(1996)\citenamefont {Perdew},
  \citenamefont {Burke},\ and\ \citenamefont {Ernzerhof}}]{pbe96}%
  \BibitemOpen
  \bibfield  {author} {\bibinfo {author} {\bibfnamefont {J.~P.}\ \bibnamefont
  {Perdew}}, \bibinfo {author} {\bibfnamefont {K.}~\bibnamefont {Burke}}, \
  and\ \bibinfo {author} {\bibfnamefont {M.}~\bibnamefont {Ernzerhof}},\
  }\bibfield  {title} {\enquote {\bibinfo {title} {Generalized gradient
  approximation made simple},}\ }\href {\doibase 10.1103/PhysRevLett.77.3865}
  {\bibfield  {journal} {\bibinfo  {journal} {Phys. Rev. Lett.}\ }\textbf
  {\bibinfo {volume} {77}},\ \bibinfo {pages} {3865--3868} (\bibinfo {year}
  {1996})}\BibitemShut {NoStop}%
\bibitem [{\citenamefont {Koepernik}\ and\ \citenamefont
  {Eschrig}(1999)}]{fplo}%
  \BibitemOpen
  \bibfield  {author} {\bibinfo {author} {\bibfnamefont {K.}~\bibnamefont
  {Koepernik}}\ and\ \bibinfo {author} {\bibfnamefont {H.}~\bibnamefont
  {Eschrig}},\ }\bibfield  {title} {\enquote {\bibinfo {title} {Full-potential
  nonorthogonal local-orbital minimum-basis band-structure scheme},}\ }\href
  {\doibase 10.1103/PhysRevB.59.1743} {\bibfield  {journal} {\bibinfo
  {journal} {Phys. Rev. B}\ }\textbf {\bibinfo {volume} {59}},\ \bibinfo
  {pages} {1743--1757} (\bibinfo {year} {1999})}\BibitemShut {NoStop}%
\bibitem [{\citenamefont {Kresse}\ and\ \citenamefont
  {Furthm\"uller}(1996{\natexlab{a}})}]{vasp1}%
  \BibitemOpen
  \bibfield  {author} {\bibinfo {author} {\bibfnamefont {G.}~\bibnamefont
  {Kresse}}\ and\ \bibinfo {author} {\bibfnamefont {J.}~\bibnamefont
  {Furthm\"uller}},\ }\bibfield  {title} {\enquote {\bibinfo {title}
  {Efficiency of ab-initio total energy calculations for metals and
  semiconductors using a plane-wave basis set},}\ }\href {\doibase
  http://dx.doi.org/10.1016/0927-0256(96)00008-0} {\bibfield  {journal}
  {\bibinfo  {journal} {Computational Materials Science}\ }\textbf {\bibinfo
  {volume} {6}},\ \bibinfo {pages} {15 -- 50} (\bibinfo {year}
  {1996}{\natexlab{a}})}\BibitemShut {NoStop}%
\bibitem [{\citenamefont {Kresse}\ and\ \citenamefont
  {Furthm\"uller}(1996{\natexlab{b}})}]{vasp2}%
  \BibitemOpen
  \bibfield  {author} {\bibinfo {author} {\bibfnamefont {G.}~\bibnamefont
  {Kresse}}\ and\ \bibinfo {author} {\bibfnamefont {J.}~\bibnamefont
  {Furthm\"uller}},\ }\bibfield  {title} {\enquote {\bibinfo {title} {Efficient
  iterative schemes for \textit{ab initio} total-energy calculations using a
  plane-wave basis set},}\ }\href {\doibase 10.1103/PhysRevB.54.11169}
  {\bibfield  {journal} {\bibinfo  {journal} {Phys. Rev. B}\ }\textbf {\bibinfo
  {volume} {54}},\ \bibinfo {pages} {11169--11186} (\bibinfo {year}
  {1996}{\natexlab{b}})}\BibitemShut {NoStop}%
\bibitem [{\citenamefont {Xiang}\ \emph {et~al.}(2013)\citenamefont {Xiang},
  \citenamefont {Lee}, \citenamefont {Koo}, \citenamefont {Gong},\ and\
  \citenamefont {Whangbo}}]{xiang2013}%
  \BibitemOpen
  \bibfield  {author} {\bibinfo {author} {\bibfnamefont {H.}~\bibnamefont
  {Xiang}}, \bibinfo {author} {\bibfnamefont {C.}~\bibnamefont {Lee}}, \bibinfo
  {author} {\bibfnamefont {H.-J.}\ \bibnamefont {Koo}}, \bibinfo {author}
  {\bibfnamefont {X.}~\bibnamefont {Gong}}, \ and\ \bibinfo {author}
  {\bibfnamefont {M.-H.}\ \bibnamefont {Whangbo}},\ }\bibfield  {title}
  {\enquote {\bibinfo {title} {Magnetic properties and energy-mapping
  analysis},}\ }\href@noop {} {\bibfield  {journal} {\bibinfo  {journal}
  {Dalton Trans.}\ }\textbf {\bibinfo {volume} {42}},\ \bibinfo {pages}
  {823--853} (\bibinfo {year} {2013})}\BibitemShut {NoStop}%
\bibitem [{SM()}]{SM}%
  \BibitemOpen
  \href@noop {} {}\bibinfo {note} {See Supplemental Material at {\cbl [\ldots]}
  for: i) neutron diffraction patterns and their refinements; ii) details of
  electronic structure calculations; iii) details on the derivation of the
  phase diagram presented in Fig.~\ref{fig:PD}; iv) theoretical evidence for
  the qualitatively different $T$-dependence of the Fe3 and Fe1/Fe2 moments.
  The Supplemental Material includes
  Refs.~\onlinecite{loop,alps,tsirlin2010,koehler1960,mcqueeney2008,abbas1983}.}\BibitemShut
  {Stop}%
\bibitem [{\citenamefont {Lyons}\ and\ \citenamefont
  {Kaplan}(1960)}]{LyonsKaplan}%
  \BibitemOpen
  \bibfield  {author} {\bibinfo {author} {\bibfnamefont {D.~H.}\ \bibnamefont
  {Lyons}}\ and\ \bibinfo {author} {\bibfnamefont {T.~A.}\ \bibnamefont
  {Kaplan}},\ }\bibfield  {title} {\enquote {\bibinfo {title} {Method for
  determining ground-state spin configurations},}\ }\href {\doibase
  10.1103/PhysRev.120.1580} {\bibfield  {journal} {\bibinfo  {journal} {Phys.
  Rev.}\ }\textbf {\bibinfo {volume} {120}},\ \bibinfo {pages} {1580--1585}
  (\bibinfo {year} {1960})}\BibitemShut {NoStop}%
\bibitem [{\citenamefont {Kaplan}\ and\ \citenamefont {Menyuk}(2007)}]{Kaplan}%
  \BibitemOpen
  \bibfield  {author} {\bibinfo {author} {\bibfnamefont {T.~A.}\ \bibnamefont
  {Kaplan}}\ and\ \bibinfo {author} {\bibfnamefont {N.}~\bibnamefont
  {Menyuk}},\ }\bibfield  {title} {\enquote {\bibinfo {title} {Spin ordering in
  three-dimensional crystals with strong competing exchange interactions},}\
  }\href {\doibase 10.1080/14786430601080229} {\bibfield  {journal} {\bibinfo
  {journal} {Phil. Mag.}\ }\textbf {\bibinfo {volume} {87}},\ \bibinfo {pages}
  {3711--2785} (\bibinfo {year} {2007})}\BibitemShut {NoStop}%
\bibitem [{\citenamefont {Luttinger}\ and\ \citenamefont {Tisza}(1946)}]{LT}%
  \BibitemOpen
  \bibfield  {author} {\bibinfo {author} {\bibfnamefont {J.~M.}\ \bibnamefont
  {Luttinger}}\ and\ \bibinfo {author} {\bibfnamefont {L.}~\bibnamefont
  {Tisza}},\ }\bibfield  {title} {\enquote {\bibinfo {title} {Theory of dipole
  interaction in crystals},}\ }\href {\doibase 10.1103/PhysRev.70.954}
  {\bibfield  {journal} {\bibinfo  {journal} {Phys. Rev.}\ }\textbf {\bibinfo
  {volume} {70}},\ \bibinfo {pages} {954--964} (\bibinfo {year}
  {1946})}\BibitemShut {NoStop}%
\bibitem [{\citenamefont {Rousochatzakis}\ \emph {et~al.}(2015)\citenamefont
  {Rousochatzakis}, \citenamefont {Richter}, \citenamefont {Zinke},\ and\
  \citenamefont {Tsirlin}}]{rousochatzakis2015}%
  \BibitemOpen
  \bibfield  {author} {\bibinfo {author} {\bibfnamefont {I.}~\bibnamefont
  {Rousochatzakis}}, \bibinfo {author} {\bibfnamefont {J.}~\bibnamefont
  {Richter}}, \bibinfo {author} {\bibfnamefont {R.}~\bibnamefont {Zinke}}, \
  and\ \bibinfo {author} {\bibfnamefont {A.~A.}\ \bibnamefont {Tsirlin}},\
  }\bibfield  {title} {\enquote {\bibinfo {title} {Frustration and
  {Dzyaloshinsky-Moriya} anisotropy in the kagome francisites
  {Cu$_3$Bi(SeO$_3)_2$O$_2$X (X = Br, Cl)}},}\ }\href {\doibase
  10.1103/PhysRevB.91.024416} {\bibfield  {journal} {\bibinfo  {journal} {Phys.
  Rev. B}\ }\textbf {\bibinfo {volume} {91}},\ \bibinfo {pages} {024416}
  (\bibinfo {year} {2015})}\BibitemShut {NoStop}%
\bibitem [{Note3()}]{Note3}%
  \BibitemOpen
  \bibinfo {note} {For the sake of comparison with Fig.~\ref {fig:SIA}, we
  provide the overall coupling energy that is \protect \textit {not} normalized
  to $S=\protect \frac 52$.}\BibitemShut {Stop}%
\bibitem [{\citenamefont {Cumby}\ \emph {et~al.}(2016)\citenamefont {Cumby},
  \citenamefont {Bayliss}, \citenamefont {Berry},\ and\ \citenamefont
  {Greaves}}]{cumby2016}%
  \BibitemOpen
  \bibfield  {author} {\bibinfo {author} {\bibfnamefont {J.}~\bibnamefont
  {Cumby}}, \bibinfo {author} {\bibfnamefont {R.~D.}\ \bibnamefont {Bayliss}},
  \bibinfo {author} {\bibfnamefont {F.~J.}\ \bibnamefont {Berry}}, \ and\
  \bibinfo {author} {\bibfnamefont {C.}~\bibnamefont {Greaves}},\ }\bibfield
  {title} {\enquote {\bibinfo {title} {Synthetic analogues of {Fe(II)--Fe(III)}
  minerals containing a pentagonal {'Cairo'} magnetic lattice},}\ }\href
  {\doibase 10.1039/c6dt01672c} {\bibfield  {journal} {\bibinfo  {journal}
  {Dalton Trans.}\ }\textbf {\bibinfo {volume} {45}},\ \bibinfo {pages}
  {11801--11806} (\bibinfo {year} {2016})}\BibitemShut {NoStop}%
\bibitem [{Note4()}]{Note4}%
  \BibitemOpen
  \bibinfo {note} {The typical single-ion anisotropy term for octahedrally
  coordinated Mn$^{3+}$ is $2-3$\protect \tmspace +\thinmuskip {.1667em}K~\cite
  {moussa1996,kida1973} yielding the energy difference on the order of
  10\protect \tmspace +\thinmuskip {.1667em}K upon spin rotation, the energy
  scale similar to that of Fe3 in Fig.~\ref {fig:SIA}.}\BibitemShut {Stop}%
\bibitem [{\citenamefont {Todo}\ and\ \citenamefont {Kato}(2001)}]{loop}%
  \BibitemOpen
  \bibfield  {author} {\bibinfo {author} {\bibfnamefont {S.}~\bibnamefont
  {Todo}}\ and\ \bibinfo {author} {\bibfnamefont {K.}~\bibnamefont {Kato}},\
  }\bibfield  {title} {\enquote {\bibinfo {title} {Cluster algorithms for
  general- $\mathit{S}$ quantum spin systems},}\ }\href {\doibase
  10.1103/PhysRevLett.87.047203} {\bibfield  {journal} {\bibinfo  {journal}
  {Phys. Rev. Lett.}\ }\textbf {\bibinfo {volume} {87}},\ \bibinfo {pages}
  {047203} (\bibinfo {year} {2001})}\BibitemShut {NoStop}%
\bibitem [{\citenamefont {Albuquerque}\ \emph {et~al.}(2007)\citenamefont
  {Albuquerque}, \citenamefont {Alet}, \citenamefont {Corboz}, \citenamefont
  {Dayal}, \citenamefont {Feiguin}, \citenamefont {Fuchs}, \citenamefont
  {Gamper}, \citenamefont {Gull}, \citenamefont {G\"urtler}, \citenamefont
  {Honecker}, \citenamefont {Igarashi}, \citenamefont {K\"orner}, \citenamefont
  {Kozhevnikov}, \citenamefont {L\"auchli}, \citenamefont {Manmana},
  \citenamefont {Matsumoto}, \citenamefont {McCulloch}, \citenamefont {Michel},
  \citenamefont {Noack}, \citenamefont {Paw{\l}owski}, \citenamefont {Pollet},
  \citenamefont {Pruschke}, \citenamefont {Schollw\"ock}, \citenamefont {Todo},
  \citenamefont {Trebst}, \citenamefont {Troyer}, \citenamefont {Werner},\ and\
  \citenamefont {Wessel}}]{alps}%
  \BibitemOpen
  \bibfield  {author} {\bibinfo {author} {\bibfnamefont {A.F.}\ \bibnamefont
  {Albuquerque}}, \bibinfo {author} {\bibfnamefont {F.}~\bibnamefont {Alet}},
  \bibinfo {author} {\bibfnamefont {P.}~\bibnamefont {Corboz}}, \bibinfo
  {author} {\bibfnamefont {P.}~\bibnamefont {Dayal}}, \bibinfo {author}
  {\bibfnamefont {A.}~\bibnamefont {Feiguin}}, \bibinfo {author} {\bibfnamefont
  {S.}~\bibnamefont {Fuchs}}, \bibinfo {author} {\bibfnamefont
  {L.}~\bibnamefont {Gamper}}, \bibinfo {author} {\bibfnamefont
  {E.}~\bibnamefont {Gull}}, \bibinfo {author} {\bibfnamefont {S.}~\bibnamefont
  {G\"urtler}}, \bibinfo {author} {\bibfnamefont {A.}~\bibnamefont {Honecker}},
  \bibinfo {author} {\bibfnamefont {R.}~\bibnamefont {Igarashi}}, \bibinfo
  {author} {\bibfnamefont {M.}~\bibnamefont {K\"orner}}, \bibinfo {author}
  {\bibfnamefont {A.}~\bibnamefont {Kozhevnikov}}, \bibinfo {author}
  {\bibfnamefont {A.}~\bibnamefont {L\"auchli}}, \bibinfo {author}
  {\bibfnamefont {S.R.}\ \bibnamefont {Manmana}}, \bibinfo {author}
  {\bibfnamefont {M.}~\bibnamefont {Matsumoto}}, \bibinfo {author}
  {\bibfnamefont {I.P.}\ \bibnamefont {McCulloch}}, \bibinfo {author}
  {\bibfnamefont {F.}~\bibnamefont {Michel}}, \bibinfo {author} {\bibfnamefont
  {R.M.}\ \bibnamefont {Noack}}, \bibinfo {author} {\bibfnamefont
  {G.}~\bibnamefont {Paw{\l}owski}}, \bibinfo {author} {\bibfnamefont
  {L.}~\bibnamefont {Pollet}}, \bibinfo {author} {\bibfnamefont
  {T.}~\bibnamefont {Pruschke}}, \bibinfo {author} {\bibfnamefont
  {U.}~\bibnamefont {Schollw\"ock}}, \bibinfo {author} {\bibfnamefont
  {S.}~\bibnamefont {Todo}}, \bibinfo {author} {\bibfnamefont {S.}~\bibnamefont
  {Trebst}}, \bibinfo {author} {\bibfnamefont {M.}~\bibnamefont {Troyer}},
  \bibinfo {author} {\bibfnamefont {P.}~\bibnamefont {Werner}}, \ and\ \bibinfo
  {author} {\bibfnamefont {S.}~\bibnamefont {Wessel}},\ }\bibfield  {title}
  {\enquote {\bibinfo {title} {{The ALPS project release 1.3: Open-source
  software for strongly correlated systems}},}\ }\href@noop {} {\bibfield
  {journal} {\bibinfo  {journal} {J. Magn. Magn. Mater.}\ }\textbf {\bibinfo
  {volume} {310}},\ \bibinfo {pages} {1187--1193} (\bibinfo {year}
  {2007})}\BibitemShut {NoStop}%
\bibitem [{\citenamefont {Tsirlin}\ \emph {et~al.}(2010)\citenamefont
  {Tsirlin}, \citenamefont {Janson},\ and\ \citenamefont
  {Rosner}}]{tsirlin2010}%
  \BibitemOpen
  \bibfield  {author} {\bibinfo {author} {\bibfnamefont {A.~A.}\ \bibnamefont
  {Tsirlin}}, \bibinfo {author} {\bibfnamefont {O.}~\bibnamefont {Janson}}, \
  and\ \bibinfo {author} {\bibfnamefont {H.}~\bibnamefont {Rosner}},\
  }\bibfield  {title} {\enquote {\bibinfo {title} {{$\beta$-Cu$_2$V$_2$O$_7$}:
  A spin-1/2 honeycomb lattice system},}\ }\href@noop {} {\bibfield  {journal}
  {\bibinfo  {journal} {Phys. Rev. B}\ }\textbf {\bibinfo {volume} {82}},\
  \bibinfo {pages} {144416} (\bibinfo {year} {2010})}\BibitemShut {NoStop}%
\bibitem [{\citenamefont {Koehler}\ \emph {et~al.}(1960)\citenamefont
  {Koehler}, \citenamefont {Wollan},\ and\ \citenamefont
  {Wilkinson}}]{koehler1960}%
  \BibitemOpen
  \bibfield  {author} {\bibinfo {author} {\bibfnamefont {W.~C.}\ \bibnamefont
  {Koehler}}, \bibinfo {author} {\bibfnamefont {E.~O.}\ \bibnamefont {Wollan}},
  \ and\ \bibinfo {author} {\bibfnamefont {M.~K.}\ \bibnamefont {Wilkinson}},\
  }\bibfield  {title} {\enquote {\bibinfo {title} {Neutron diffraction study of
  the magnetic properties of rare-earth-iron perovskites},}\ }\href {\doibase
  10.1103/PhysRev.118.58} {\bibfield  {journal} {\bibinfo  {journal} {Phys.
  Rev.}\ }\textbf {\bibinfo {volume} {118}},\ \bibinfo {pages} {58--70}
  (\bibinfo {year} {1960})}\BibitemShut {NoStop}%
\bibitem [{\citenamefont {McQueeney}\ \emph {et~al.}(2008)\citenamefont
  {McQueeney}, \citenamefont {Yan}, \citenamefont {Chang},\ and\ \citenamefont
  {Ma}}]{mcqueeney2008}%
  \BibitemOpen
  \bibfield  {author} {\bibinfo {author} {\bibfnamefont {R.~J.}\ \bibnamefont
  {McQueeney}}, \bibinfo {author} {\bibfnamefont {J.-Q.}\ \bibnamefont {Yan}},
  \bibinfo {author} {\bibfnamefont {S.}~\bibnamefont {Chang}}, \ and\ \bibinfo
  {author} {\bibfnamefont {J.}~\bibnamefont {Ma}},\ }\bibfield  {title}
  {\enquote {\bibinfo {title} {Determination of the exchange anisotropy in
  perovskite antiferromagnets using powder inelastic neutron scattering},}\
  }\href {\doibase 10.1103/PhysRevB.78.184417} {\bibfield  {journal} {\bibinfo
  {journal} {Phys. Rev. B}\ }\textbf {\bibinfo {volume} {78}},\ \bibinfo
  {pages} {184417} (\bibinfo {year} {2008})}\BibitemShut {NoStop}%
\bibitem [{\citenamefont {Abbas}\ \emph {et~al.}(1983)\citenamefont {Abbas},
  \citenamefont {Mostafa},\ and\ \citenamefont {Fayek}}]{abbas1983}%
  \BibitemOpen
  \bibfield  {author} {\bibinfo {author} {\bibfnamefont {Y.}~\bibnamefont
  {Abbas}}, \bibinfo {author} {\bibfnamefont {F.}~\bibnamefont {Mostafa}}, \
  and\ \bibinfo {author} {\bibfnamefont {M.}~\bibnamefont {Fayek}},\ }\bibfield
   {title} {\enquote {\bibinfo {title} {Antiferromagnetic structure of barium
  strontium {tetraferrate(III)}, {BaSrFe$_4$O$_8$}},}\ }\href@noop {}
  {\bibfield  {journal} {\bibinfo  {journal} {Acta Cryst.}\ }\textbf {\bibinfo
  {volume} {39}},\ \bibinfo {pages} {1--4} (\bibinfo {year}
  {1983})}\BibitemShut {NoStop}%
\bibitem [{\citenamefont {Moussa}\ \emph {et~al.}(1996)\citenamefont {Moussa},
  \citenamefont {Hennion}, \citenamefont {Rodriguez-Carvajal}, \citenamefont
  {Moudden}, \citenamefont {Pinsard},\ and\ \citenamefont
  {Revcolevschi}}]{moussa1996}%
  \BibitemOpen
  \bibfield  {author} {\bibinfo {author} {\bibfnamefont {F.}~\bibnamefont
  {Moussa}}, \bibinfo {author} {\bibfnamefont {M.}~\bibnamefont {Hennion}},
  \bibinfo {author} {\bibfnamefont {J.}~\bibnamefont {Rodriguez-Carvajal}},
  \bibinfo {author} {\bibfnamefont {H.}~\bibnamefont {Moudden}}, \bibinfo
  {author} {\bibfnamefont {L.}~\bibnamefont {Pinsard}}, \ and\ \bibinfo
  {author} {\bibfnamefont {A.}~\bibnamefont {Revcolevschi}},\ }\bibfield
  {title} {\enquote {\bibinfo {title} {Spin waves in the antiferromagnet
  perovskite {LaMnO$_3$}: A neutron-scattering study},}\ }\href {\doibase
  10.1103/PhysRevB.54.15149} {\bibfield  {journal} {\bibinfo  {journal} {Phys.
  Rev. B}\ }\textbf {\bibinfo {volume} {54}},\ \bibinfo {pages} {15149--15155}
  (\bibinfo {year} {1996})}\BibitemShut {NoStop}%
\bibitem [{\citenamefont {Kida}\ and\ \citenamefont
  {Watanabe}(1973)}]{kida1973}%
  \BibitemOpen
  \bibfield  {author} {\bibinfo {author} {\bibfnamefont {J.}~\bibnamefont
  {Kida}}\ and\ \bibinfo {author} {\bibfnamefont {T.}~\bibnamefont
  {Watanabe}},\ }\bibfield  {title} {\enquote {\bibinfo {title} {Anisotropy and
  weak ferromagnetism in linear chain antiferromagnet {(NH$_4)_2$MnF$_5$}},}\
  }\href {\doibase 10.1143/JPSJ.34.952} {\bibfield  {journal} {\bibinfo
  {journal} {J. Phys. Soc. Jpn.}\ }\textbf {\bibinfo {volume} {34}},\ \bibinfo
  {pages} {952--958} (\bibinfo {year} {1973})}\BibitemShut {NoStop}%
\end{thebibliography}
%

\clearpage
\renewcommand{\thefigure}{S\arabic{figure}}
\renewcommand{\thetable}{S\arabic{table}}
\renewcommand{\thesection}{S\arabic{section}}
\setcounter{figure}{0}
\setcounter{table}{0}
\setcounter{section}{0}

\begin{widetext}
\begin{center}
{\large\bf Supplemental Material}
\end{center}
\end{widetext}

\begin{widetext}
\begin{table*}[!h]
\caption{\label{tab:refinement}
Refined magnetic moments at different temperatures. The vectors of magnetic moments are represented using spherical coordinates $(\mu,\theta,\varphi)$, where $\mu$ is the magnetic moment value and $\theta$ and $\varphi$ are the azimuthal and polar angles, respectively. The magnetic moment components are then expressed as follows: $m_a=\mu\sin\theta\cos\varphi$, $m_b=\mu\sin\theta\sin\varphi$, and $m_c=\mu\cos\theta$. The angle $\theta$ is fixed to $90^{\circ}$ (all spins are in the $ab$ plane), whereas $\varphi$ is measured with respect to the $\mathbf a_m$ direction. 
}
\begin{ruledtabular}
\begin{tabular}{cccccccccccc}
 $T$ & $\mu$(Fe1$_1$) & $\varphi$(Fe1$_1$) & $\mu$(Fe1$_2$) & $\varphi$(Fe1$_2$) & $\mu$(Fe2$_1$) & $\varphi$(Fe2$_1$) & $\mu$(Fe2$_2$) & $\varphi$(Fe2$_2$) & $\mu$(Fe3) & $\varphi$(Fe3) & $R_{\rm mag}$ \\\hline

1.5 & 3.92(2) & 86.1(5) & $\mu$(Fe1$_1$) & $90^{\circ}\!+\!\varphi$(Fe1$_1$) & 3.33(3) & 52.4(9) & $\mu$(Fe2$_1$) & $-90^{\circ}\!+\!\varphi$(Fe2$_1$) & $\mu$(Fe1$_1$) & $-180^{\circ}\!+\!\varphi$(Fe1$_1$) & 0.020 \\

30  & 3.82(4) & 83.2(7) & $\mu$(Fe1$_1$) & $90^{\circ}\!+\!\varphi$(Fe1$_1$) & 3.33(4) & 52(1)   & $\mu$(Fe2$_1$) & $-90^{\circ}\!+\!\varphi$(Fe2$_1$) & 3.02(3)        & $-180^{\circ}\!+\!\varphi$(Fe1$_1$) & 0.025 \\

50  & 3.63(4) & 80.2(8) & $\mu$(Fe1$_1$) & $90^{\circ}\!+\!\varphi$(Fe1$_1$) & 3.19(4) & 48(1)   & $\mu$(Fe2$_1$) & $-90^{\circ}\!+\!\varphi$(Fe2$_1$) & 2.10(3)        & $-180^{\circ}\!+\!\varphi$(Fe1$_1$) & 0.026 \\

55  & 3.49(3) & 78.8(6) & $\mu$(Fe1$_1$) & $90^{\circ}\!+\!\varphi$(Fe1$_1$) & 3.14(3) & 48(1)   & $\mu$(Fe2$_1$) & $-90^{\circ}\!+\!\varphi$(Fe2$_1$) & 1.91(2)        & $-180^{\circ}\!+\!\varphi$(Fe1$_1$) & 0.025 \\

60  & 3.44(5) & 94.0(9) & 3.56(5)    & 	155.3(9)                       & 2.31(7)   & 48(1)   & $\mu$(Fe1$_1$) & $-48(2)$                    & 1.69(2)        & $-102(2)$                       & 0.022 \\

62.5& 3.73(4) & 108(1)  & 3.21(8)    & 	129(1)                         & 1.1(1)    & 76(4)   & $\mu$(Fe1$_1$) & $-69(2)$                    & 1.32(3)        & $-124(4)$                       & 0.028 \\

65  & 3.70(3) & 112(1)  & 3.07(6)    & 	120(1)                         & 1.1(1)    & 98(2)   & $\mu$(Fe1$_1$) & $-75(2)$                    & 1.15(3)        & $-128(4)$                       & 0.028 \\

67.5& 3.77(4) & 114(1)  & 3.14(7)    & 	116(1)                         & 1.1(1)    & 105(2)  & $\mu$(Fe1$_1$) & $-77(2)$                    & 1.09(5)        & $-140(6)$                       & 0.030 \\

70  & 3.64(3) & 119(1)  & 3.10(6)    & 	116(1)                         & 1.0(1)    & 119(3)  & $\mu$(Fe1$_1$) & $-77(2)$                    & 1.01(4)        & $-136(5)$                       & 0.028 \\

75  & 3.88(6) & 111(1)  & 1.89(7)    & $-90^{\circ}\!+\!\varphi$(Fe1$_1$)  & 2.42(5)   & 139(3)  & $\mu$(Fe2$_1$) & $-116(3)$                    & 0.4(2)         & $-179(9)$                       & 0.035 \\

80  & 3.84(5) & 116(1)  & 2.21(5)    & $-90^{\circ}\!+\!\varphi$(Fe1$_1$)  & 2.60(3)   & 147(2)  & $\mu$(Fe2$_1$) & $-270^{\circ}\!+\!\varphi$(Fe2$_1$)  & 0        & --                              & 0.038 \\

90  & 3.83(6) & 117(1)  & 2.32(7)    & $-90^{\circ}\!+\!\varphi$(Fe1$_1$)  & 2.66(4)   & 149(2)  & $\mu$(Fe2$_1$) & $-270^{\circ}\!+\!\varphi$(Fe2$_1$)  & 0        & --                              & 0.031 \\

100 & 3.73(6) & 116(1)  & 2.34(7)    & $-90^{\circ}\!+\!\varphi$(Fe1$_1$)  & 2.66(4)   & 147(2)  & $\mu$(Fe2$_1$) & $-270^{\circ}\!+\!\varphi$(Fe2$_1$)  & 0        & --                              & 0.029 \\

110 & 3.58(9) & 119(1)  & 2.52(9)    & $-90^{\circ}\!+\!\varphi$(Fe1$_1$)  & 2.71(4)   & 149(2)  & $\mu$(Fe2$_1$) & $-270^{\circ}\!+\!\varphi$(Fe2$_1$)  & 0        & --                              & 0.030 \\

120 & 3.57(6) & 120(1)  & 2.44(6)    & $-90^{\circ}\!+\!\varphi$(Fe1$_1$)  & 2.68(3)   & 149(1)  & $\mu$(Fe2$_1$) & $-270^{\circ}\!+\!\varphi$(Fe2$_1$)  & 0        & --                              & 0.029 \\

140 & 3.22(6) & 119(1)  & 2.11(6)    & $-90^{\circ}\!+\!\varphi$(Fe1$_1$)  & 2.44(3)   & 148(1)  & $\mu$(Fe2$_1$) & $-270^{\circ}\!+\!\varphi$(Fe2$_1$)  & 0        & --                              & 0.033 \\

160 & 2.71(6) & 116(1)  & 1.80(6)    & $-90^{\circ}\!+\!\varphi$(Fe1$_1$)  & 1.98(3)   & 149(2)  & $\mu$(Fe2$_1$) & $-270^{\circ}\!+\!\varphi$(Fe2$_1$)  & 0        & --                              & 0.035 \\
\end{tabular}
\end{ruledtabular}
\end{table*}

\begin{figure*}[!h]
\includegraphics[width=0.999\textwidth,angle=0,clip=true,trim=0 0 0 0]{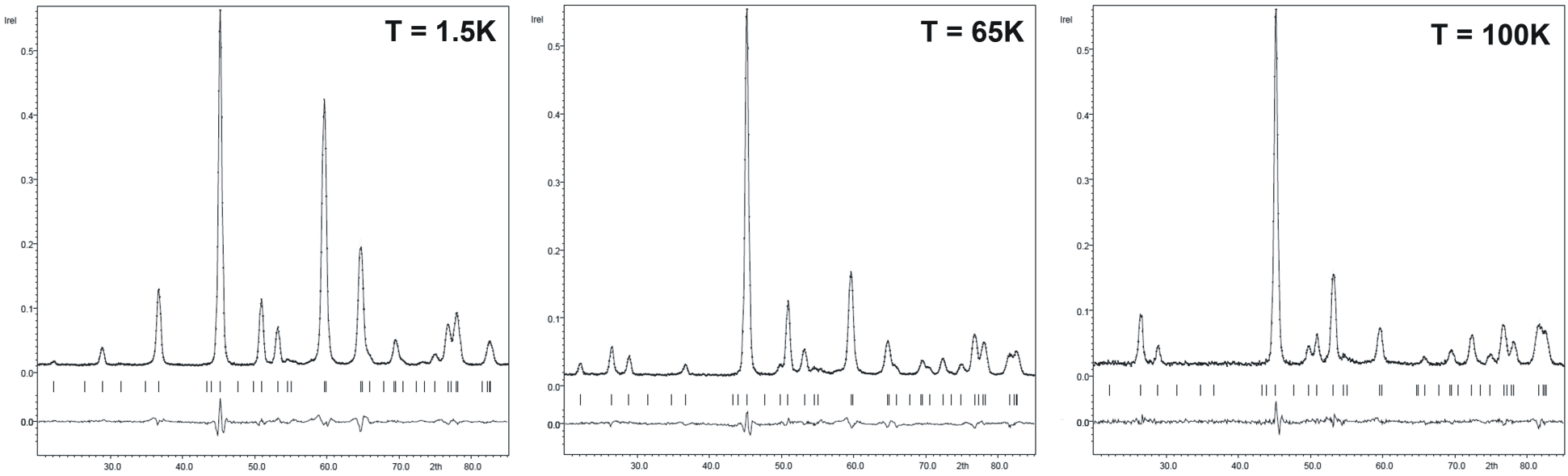}
\caption{\label{fig:patterns} Refined neutron diffraction patterns at 1.5\,K, 65\,K, and 100\,K.}
\end{figure*}

\begin{figure*}[!h]
\includegraphics[width=0.75\textwidth,angle=0,clip=true,trim=0 0 0 0]{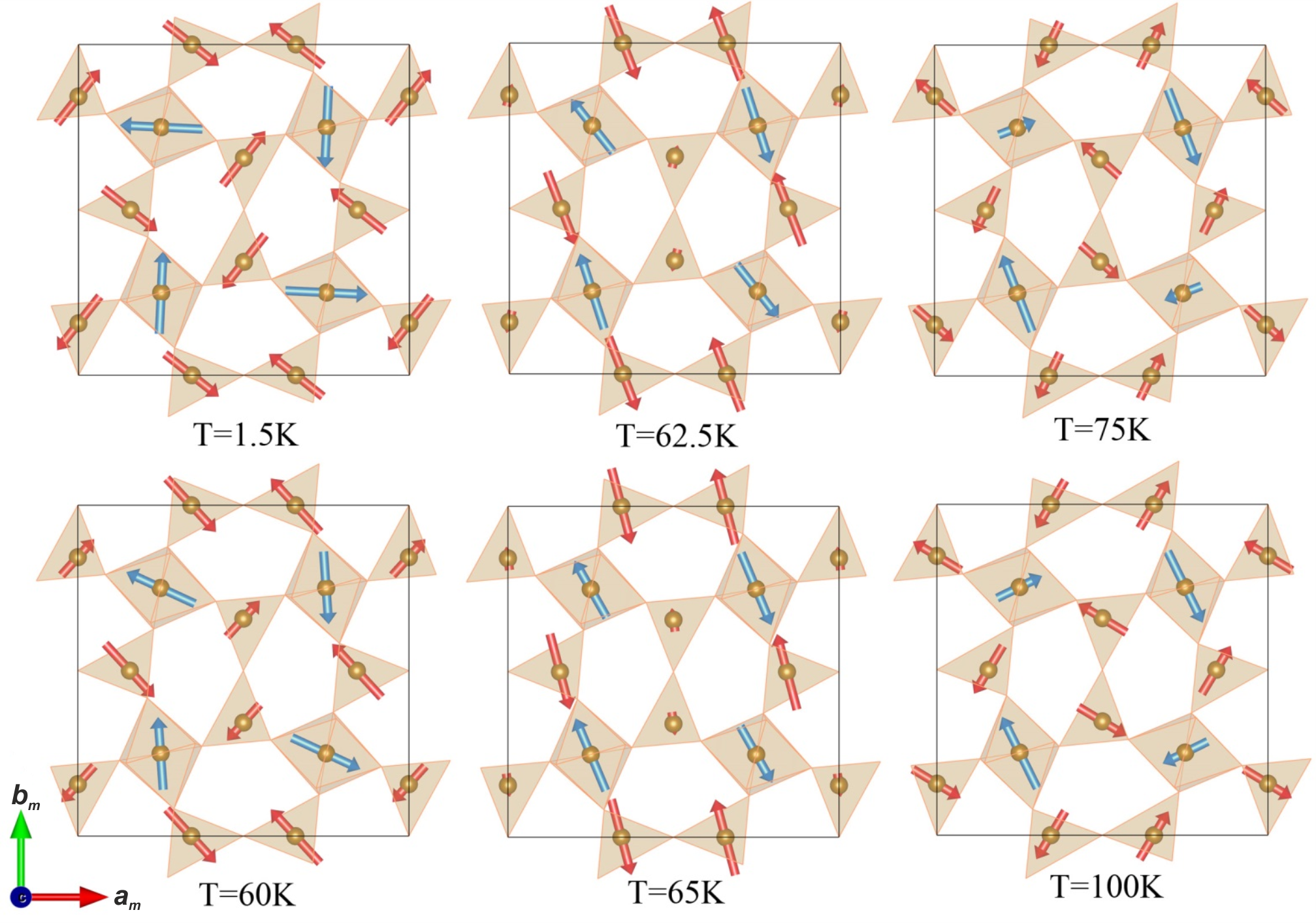}
\caption{\label{fig:magstructures} Temperature evolution of the magnetic order in the Cairo plane. The Cairo planes at \textit{z} = 0 are shown.}
\end{figure*}
\end{widetext}
\clearpage

\section{Isotropic exchange couplings}
Isotropic exchange couplings were obtained from total energies of collinear and orthogonal spin configurations using the mapping procedure~\cite{xiang2013}. We chose the Perdew-Burke-Ernzerhof (PBE) flavor of the exchange-correlation potential~\cite{pbe96}, and included correlations on the mean-field DFT+$U$ level. Different values of the on-site Coulomb repulsion parameter $U_d$ were applied for the octahedrally and tetrahedrally coordinated Fe sites. The choice of $U_d$ was justified by calculating magnetic exchange couplings in reference compounds LaFeO$_3$ (octahedrally coordinated Fe$^{3+}$ ions) and BaSrFe$_4$O$_8$ (tetrahedrally coordinated Fe$^{3+}$ ions). For both compounds, we obtained exchange couplings using different values of $U_d$, and estimated N\'eel temperatures by simulating magnetic susceptibility with the \texttt{loop} algorithm~\cite{loop} of \texttt{ALPS}~\cite{alps}. Note that the value of $U_d$ is also dependent on the band-structure code, given the different basis sets for $3d$-orbitals. The results of \texttt{FPLO} calculations can be well reproduced in \texttt{VASP} by increasing $U_d$ for 1\,eV, which is similar to our earlier experience~\cite{tsirlin2010}.

In LaFeO$_3$, the \texttt{FPLO} calculations with $U_d=6$\,eV (or, respectively, the \texttt{VASP} calculations with $U_d=7$\,eV) yield $J_{ab}=61$\,K and $J_c=54$\,K for the couplings in the $ab$ plane and along the $c$ direction, respectively, resulting in the N\'eel temperature $T_N=729$\,K. This value is in perfect agreement with the experimental $T_N\simeq 730$\,K~\cite{koehler1960}. Our computed exchange couplings also reproduce the averaged exchange coupling $\bar J=56.8$\,K obtained from inelastic neutron scattering~\cite{mcqueeney2008}. 

In BaSrFe$_4$O$_8$, honeycomb bilayers of the Fe$^{3+}$ ions are formed. Using $U_d=8$\,eV in \texttt{FPLO} or $U_d=9$\,eV in \texttt{VASP}, we obtained the in-plane coupling $J_{ab}=103$\,K, the coupling $J_{\perp}=175$\,K between the two layers of the bilayer, and the coupling $J_c=0.7$\,K between the bilayers. This set of exchange couplings yields $T_N=700$\,K in excellent agreement with the experimental value of 690\,K~\cite{abbas1983}. If, on the other hand, $U_d=6$\,eV is used (\texttt{FPLO}), $T_N$ increases well above 800\,K implying that higher values of $U_d$ are required for a proper description of the tetrahedrally coordinated Fe$^{3+}$ sites.

\begin{table}
\caption{\label{tab:exchange}
Exchange couplings in Bi$_4$Fe$_5$O$_{13}$F: the Fe--Fe distances $d$ (in\,\r A), the Fe--O--Fe angles $\psi$ (in\,deg), and exchange couplings $J_i$ (in\,K) obtained from \texttt{FPLO} calculations. The values from Ref.~[\onlinecite{abakumov2013}] were obtained using $U_d=7$\,eV and $J_d=1$\,eV. The values in the current work are based on the optimized choice of $U_d$, 6\,eV for the octahedrally coordinated Fe sites and 8\,eV for the tetrahedrally coordinated Fe sites.
}
\begin{ruledtabular}
\begin{tabular}{ccccc}
  & $d_{\rm Fe-Fe}$ & $\psi_{\rm Fe-O-Fe}$ &  \multicolumn{2}{c}{$J_i$} \\
	&  & & Ref.~[\onlinecite{abakumov2013}] & this work \\\hline
$J_{43}$  & 	3.39 & 119.2 & 45  & 38  \\
$J_{43}'$ &   3.53 & 130.9 & 74  & 57  \\
$J_{33}$  &   3.64 & 180   & 191 & 116 \\
$J_{44}$  &   2.91 & 94.2  & 34  & 9   \\
$J_{\perp}$ & 3.06 & 97.4  & 10  & 3   \\
$J_{\perp2}$ & 6.11 & -- & -- & 2   \\
\end{tabular}
\end{ruledtabular}
\end{table}

Using such optimized values of $U_d$, we calculated exchange couplings in Bi$_4$Fe$_5$O$_{13}$F. In Table~\ref{tab:exchange}, they are compared with the values obtained in Ref.~[\onlinecite{abakumov2013}], where same value of $U_d$ was applied to all Fe sites. While no qualitative changes are observed, the coupling $J_{33}$ between the two tetrahedrally coordinated Fe atoms is reduced substantially, because a higher value of $U_d$ is applied. This reduction has an immediate effect on the N\'eel temperature that equals to 250\,K for the parameter set from Ref.~[\onlinecite{abakumov2013}] and 150\,K for our new parameter set. An even better agreement with the experimental $T_N=180$\,K can be achieved by adjusting the weakest coupling $J_{\perp}$. We found that $J_{\perp}=8$\,K leads to the best agreement not only with the experimental $T_N$, but also with the magnetic susceptibility curve in general. We have also identified the second-neighbor coupling $J_{\perp 2}$ along the $c$ axis. It is responsible for the AFM interlayer order in phase III. 

\section{Magnetic anisotropy}
For calculating single-ion anisotropy, we used the same mapping approach~\cite{xiang2013} but exploited orthogonal spin configurations, where the spin of interest probes different directions in the $ab$ plane, whereas all other spins are along the $c$ direction, so that all interactions with other spins are canceled, and the effect of single-ion anisotropy is probed exclusively. While spin configurations of our choice eliminate all intersite interactions for Fe1 and Fe3, the calculations for Fe2 required two antiparallel spins on one $J_{33}$ bond to be used simultaneously, in order to eliminate contributions of Dzyaloshinsky-Moriya couplings on the bonds $J_{43}$ and $J_{43}'$. All calculations were performed on the DFT+$U$+SO level in \texttt{VASP}, hence we chose $U_d=7$\,eV for the octahedrally coordinated Fe and $U_d=9$\,eV for the tetrahedrally coordinated Fe.

\begin{table}
\caption{
Single-ion anisotropy of Fe$^{3+}$ ions parametrized using Eq.~\eqref{eq:anis}. The pre-factor $A$ is in K, and the phase shift $\varphi_0$ is in degrees, showing the position of the energy minimum (preferred spin direction) for a given atom.
}
\begin{ruledtabular}
\begin{tabular}{ccc@{\hspace{4em}}ccc}\smallskip
  & $A$ & $\varphi_0$ & & $A$ & $\varphi_0$ \\
	Fe1$_1$ & 0.75 & 110.9 & Fe2$_1$ & 0.69 & 47.5 \\
	Fe1$_2$ & 0.75 & 159.1 & Fe2$_2$ & 0.69 & 42.5 \\
	Fe3 & 4.07 & 90 & & & \\
\end{tabular}
\end{ruledtabular}
\end{table}

Total energies as a function of the polar angle $\varphi$ (Fig.~\ref{fig:SIA} of the manuscript) were parametrized as follows,
\begin{equation}
 E=A\,(1-\cos[\,2\,(\varphi-\varphi_0)\,]),
\label{eq:anis}\end{equation}
where $A$ is the magnitude of the single-ion anisotropy, and $\varphi_0$ defines preferred spin direction in the $ab$ plane. Absolute values of the anisotropy are relatively small, given the $d^5$ nature of the Fe$^{3+}$ ions. The much larger value of $A$ for Fe3 compared to Fe1 can be related to the strongly distorted nature of the Fe3O$_6$ octahedra. Remarkably, the calculated single-ion anisotropy with $\varphi_0=90^{\circ}$ for Fe3 favors model B and disfavors model A for the magnetic structure of phase I, in agreement with the experiment.

For the sake of completeness, we also compared single-ion energies for in-plane and out-of-plane spin directions and obtained 2.7\,K for Fe1, $-0.55$\,K for Fe2, and $-2.9$\,K for Fe3, where positive sign implies easy-plane anisotropy ($ab$ plane) and negative sign implies easy-axis anisotropy ($c$ direction). Although the easy-axis anisotropy of Fe3 is slightly stronger than the easy-plane anisotropy of Fe1, the easy-plane nature of Fe1 spins prevails, because the Fe1 atoms are twice more abundant compared to Fe3.

\newpage
\begin{widetext}
\section{Classical phase diagram of the model}
Figure~\ref{fig:Model} shows the topology of the Heisenberg interactions within one unit cell of \cairo. The system has a square Bravais lattice, defined by the primitive translations $\vec{t}_1$ and $\vec{t}_2$ (the ones spanning the shaded region of the figure). There are ten sites per unit cell labeled by the numbers 1-10. Among these, \{1, 2, 3, 4\} are Fe1 sites, \{5, 6, 7, 8\} are Fe2 sites, and \{9,10\} are Fe3 sites. Sites 1, 2 and 9 sit on top of each other, and the same is true for sites 3,4, and 10, see figure.
In the following we shall denote the spins by $\vec{S}_{\vec{r},\nu}$, with $\vec{r}$ labeling the primitive position of the unit cell and $\nu=1$-$10$.
The isotropic Heisenberg Hamiltonian reads: 
\small
\bea
\mc{H}&=&\sum_{\vec{r}}\Big\{
J_{44} \left(\vec{S}_{\vec{r},1}\cdot \vec{S}_{\vec{r},2} 
+\vec{S}_{\vec{r},3}\cdot \vec{S}_{\vec{r},4} \right)
+J_{\perp} \left(\vec{S}_{\vec{r},1}\cdot \vec{S}_{\vec{r}+\vec{c},9} 
+\vec{S}_{\vec{r},2}\cdot \vec{S}_{\vec{r},9} 
+\vec{S}_{\vec{r},3}\cdot \vec{S}_{\vec{r}+\vec{c},10} 
+\vec{S}_{\vec{r},4}\cdot \vec{S}_{\vec{r},10}\right)\nonumber\\
&+&
J_{43} \left[ 
\left(\vec{S}_{\vec{r},1}+\vec{S}_{\vec{r},2}\right)\cdot \left(\vec{S}_{\vec{r},5}+\vec{S}_{\vec{r},7}\right)
+\left(\vec{S}_{\vec{r},3}+\vec{S}_{\vec{r},4}\right)\cdot \left(\vec{S}_{\vec{r},6}+\vec{S}_{\vec{r}-\vec{t}_1+\vec{t}_2,8}\right)
\right]\nonumber\\
&+&
J_{43}'\left[
\left(\vec{S}_{\vec{r},1}+\vec{S}_{\vec{r},2}\right)\cdot  \left(\vec{S}_{\vec{r},6}+\vec{S}_{\vec{r},8}\right)
+\left(\vec{S}_{\vec{r},3}+\vec{S}_{\vec{r},4}\right)\cdot \left(\vec{S}_{\vec{r}-\vec{t}_1,5}+\vec{S}_{\vec{r}+\vec{t}_2,7}\right)
\right]\nonumber\\
&+&
J_{33} \left(\vec{S}_{\vec{r},5}\cdot \vec{S}_{\vec{r}+\vec{t}_2,7}
+ \vec{S}_{\vec{r},6}\cdot \vec{S}_{\vec{r}-\vec{t}_1,8}\right)
\Big\}\,,
\nonumber
\eea
\normalsize
where all couplings are antiferromagnetic (positive).

\begin{figure}
\includegraphics[width=0.4\textwidth,angle=0,clip=true,trim=0 0 0 0]{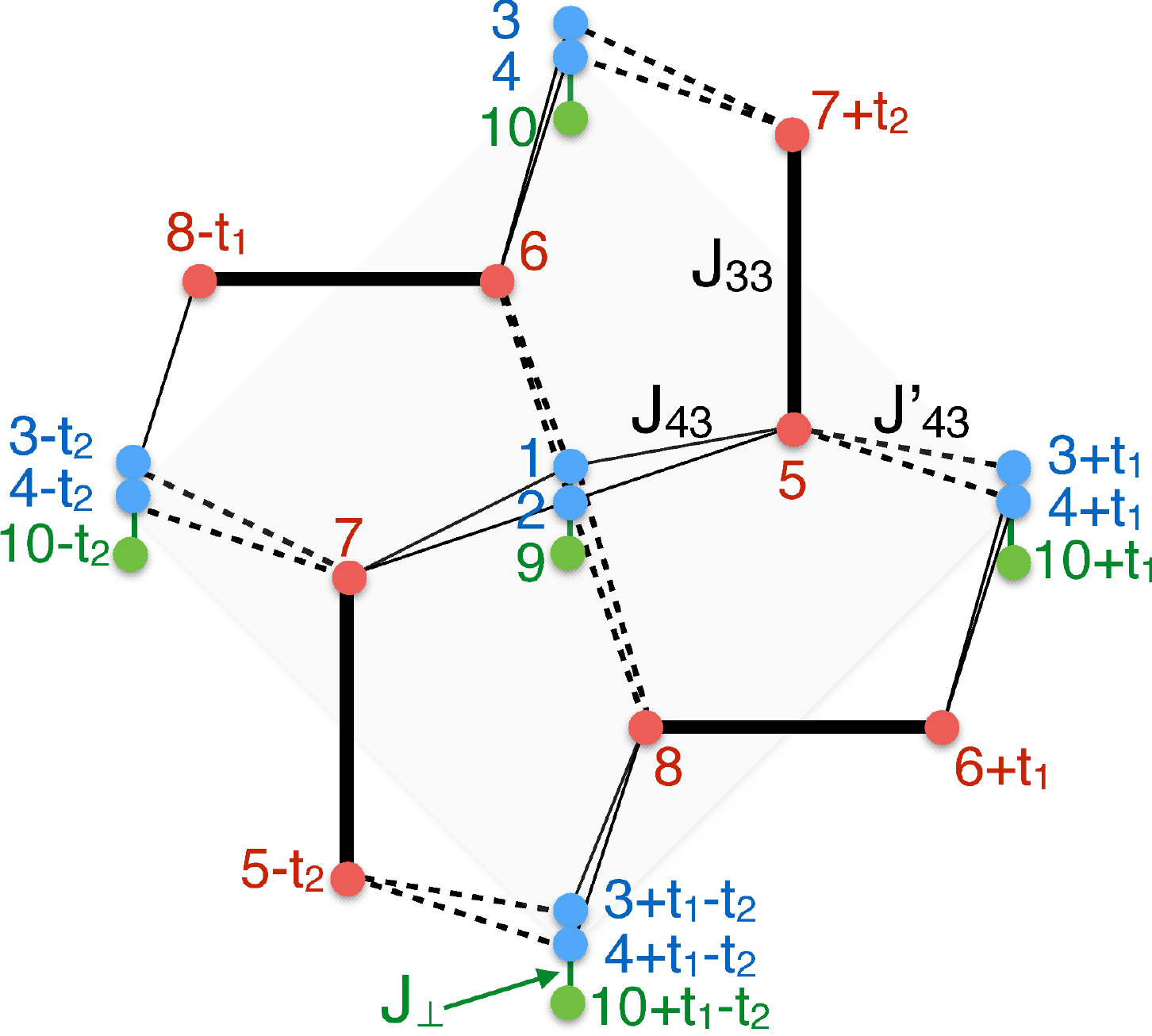}
\caption{\label{fig:Model} The unit cell of \cairo{}, along with the notation of the different sites and the connectivity of the isotropic exchange couplings. There is a pair of Fe1 sites, \{1,2\} or \{3,4\}, sitting on each of the four-fold sites of the Cairo lattice, one above and the other below the plane. Each of the spins of the pair couples with $J_{43}$ or $J_{43}'$, respectively, to the Fe2 sites with \{5,6,7,8\}. The Fe3 sites \{9,10\} sit right below and above the Fe1 sites, connecting neighboring planes to each other via $J_\perp$. The coupling $J_{44}$ (not shown) connects $1$ with $2$, and $3$ with $4$.}
\end{figure}

Before we establish the classical ground state phase diagram of the model we list a number of states that are of main interest. 

\subsection{Orthogonal states}
We begin with the two choices of orthogonal states with ordering wavevector $\vec{Q}=(\pi,\pi,0)$. 
We write: 
\small
\bea\label{fig:pipi0}
\begin{array}{l}
\vec{S}_{\vec{r},1}=\vec{S}_{\vec{r},2}=-\vec{S}_{\vec{r},9}=(-1)^{n+m} S (1,0) \\
\vec{S}_{\vec{r},3}=\vec{S}_{\vec{r},4}=-\vec{S}_{\vec{r},10}=(-1)^{n+m} S (\cos\phi_3,\sin\phi_3) \\
\vec{S}_{\vec{r},5}=\vec{S}_{\vec{r},7}=(-1)^{n+m} S (\cos\phi_5,\sin\phi_5), \\
\vec{S}_{\vec{r},6}=\vec{S}_{\vec{r},8}=(-1)^{n+m} S (\cos\phi_6,\sin\phi_6), \\
\vec{S}_{\vec{r},9}=\vec{S}_{\vec{r},4}=(-1)^{n+m} S (\cos\phi_3,\sin\phi_3), 
\end{array}
\eea
\normalsize
with 
\small
\be
\phi_6=\phi_5+\phi_3,~~
\phi_3=\chi \frac{\pi}{2},~~
\cos\phi_5=\frac{-1}{\sqrt{1+x^2}},~~
\sin\phi_5=\chi \frac{x}{\sqrt{1+x^2}},~~
x\equiv \frac{J_{43}'}{J_{43}},~~
\chi=\pm1\,.
\ee
\normalsize
The two choices of $\chi=\pm1$ correspond to the two chiralities, defined by $\bs{\chi}=\chi\vec{c}=\vec{S}_{\vec{r},1}\times\vec{S}_{\vec{r},3}/S^2$, see Fig.~\ref{fig:Phases}\,(a-b) where the two states are denoted by `orthogonal A' and `orthogonal B'. 
The strengths of the local exchange fields exerted at the mean-field level at each site are given by: 
\small
\bea
\begin{array}{l}
h_1=h_2=h_3=h_4=(J_{44}-J_{\perp}-2 |\!\cos\phi_5|J_{43}-2|\!\sin\phi_5|J_{43}')S\\
h_5=h_6=h_7=h_8=(-J_{33}-2|\!\cos\phi_5|J_{43}-2|\!\sin\phi_5|J_{43}')S\\
h_9=h_{10}=-2J_{\perp}S
\end{array}
\eea
\normalsize
and the energy of the two orthogonal solutions is given by
\small
\bea
\frac{E_\text{ortho}}{N_{uc}S^2}=2J_{44}-4J_{\perp}-2J_{33}-8|\cos\phi_5| J_{43} -8|\sin\phi_5| J_{43}' \,,
\eea
\normalsize
where $N_{uc}$ is the number of unit cells.

\subsection{Ferrimagnetic state}
Next we examine the ferrimagnetic state, which has ordering wavevector $\vec{Q}=0$. In this state, the Fe1 spins point in one direction and the Fe2 and Fe3 spins point in the opposite direction, see Fig.~\ref{fig:Phases}\,(c). Here, the local fields are given by: 
\small
\bea
\begin{array}{l}
h_1=h_2=h_3=h_4=(J_{44}-J_{\perp}-2 J_{43}- 2J_{43}')S\\
h_5=h_6=h_7=h_8=(J_{33}-2J_{43}- 2J_{43}')S\\
h_9=h_{10}=-2J_{\perp}S\,,
\end{array}
\eea
\normalsize
and the energy is 
\small
\bea
\frac{E_\text{ferri}}{N_{uc}S^2}=2J_{44}-4J_{\perp}+2J_{33}-8 (J_{43}+J_{43}')\,.
\eea
\normalsize

\begin{figure}
\includegraphics[width=0.99\textwidth,angle=0,clip=true,trim=0 0 0 0]{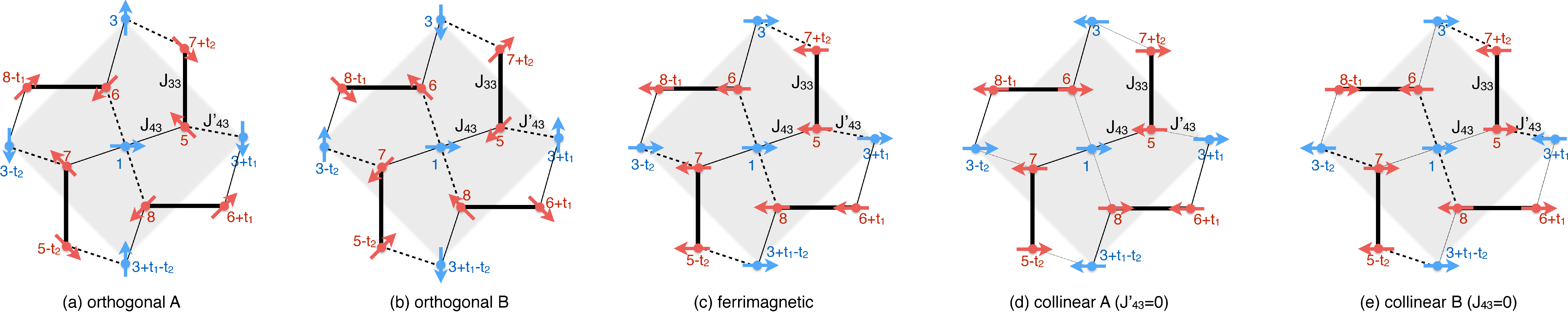}
\caption{\label{fig:Phases} Five of the classical phases discussed here. Only Fe1 and Fe2 sites are shown. (a-b) The two orthogonal configurations with opposite chiralities. (c) The ferrimagnetic state. (d) One of the two `collinear A' states appearing for $J_{43}'\!=\!0$. (e) One of the two `collinear B' states appearing for $J_{43}\!=\!0$. Note that for $J_{43}'\!=\!0$ or $J_{43}\!=\!0$ the degeneracy is actually much larger, since in these limits we get decoupled chains, and each chain can have a N\'eel order along an independent direction. The two collinear states A and B (as well as the orthogonal A and B) are special members of this highly degenerate manifold.}
\end{figure}

\subsection{Mixed state}
Let us also examine the mixed state of Ref.~[\onlinecite{rousochatzakis2012}], which has ordering wavevector $\vec{Q}=(\pi,\pi,0)$. This state combines the orthogonal state in the plane and the ferrimagnetic state perpendicular to the plane. Specifically, this state can be obtained by starting from the orthogonal state, say in the $xy$ plane,  and then tilting the spins out of the plane as follows: the spins $1,2,3,4$ tilt above the $xy$ plane by an angle $\theta$, the spins $9, 10$ tilt below the plane by the same angle $\theta$, and finally the spins $5,6,7,8$ tilt below the plane by an angle $\theta'$. The total energy can be found by combining the energies from the orthogonal and the ferrimagnetic state:
\small
\bea
\frac{ E_{\text{tot}} }{N_{uc}S^2}&=& 
2J_{44}-4J_{\perp}
-2J_{33}\cos(2\theta')-8 J_{43}\sqrt{1+x^2}\cos\theta\cos\theta'
-8(J_{43}+J_{43}') \sin\theta\sin\theta'\,.
\eea
\normalsize
Minimizing with respect to $\theta$ and $\theta'$ gives:
\small
\bea
\tan\theta' = \frac{\sqrt{1+x^2}}{1+x}\tan\theta,~~
\cos\theta'=\frac{2J_{43}J_{43}'}{J_{12}J_{33}} \cos\theta\,,
\eea
\normalsize
and combining the two equations gives
\small
\be
\sin\theta'=\frac{2J_{43}J_{43}'}{(J_{43}+J_{43}') J_{33}} \sin\theta,~~
\sin^2\theta = \frac{(J_{43}+J_{43}')^2[ (2J_{43}J_{43}')^2-J_{33}^2(J_{43}^2+J_{43}'^2)]}{(2J_{43}J_{43}')^3}\,.
\ee
\normalsize
Note that the corresponding relations for the angles $\theta$ and $\theta'$ given in Ref.~[\onlinecite{rousochatzakis2012}] for the special case of $J_{43}=J_{43}'$ can be recovered from the above relations by setting $J_{43}=J_{43}'$ and $J_{43}\to J_{43}/2$. The extra factor of two must be inserted to take into account that in the present model we have two Fe1 sites sitting at each four-fold site of the Cairo lattice, and not one spin as in Ref.~[\onlinecite{rousochatzakis2012}].

Finally, let us rewrite the energy of the mixed phase using the solutions for the angles $\theta$ and $\theta'$: 
\small
\be
\frac{E_{\text{mixed}}}{N_{uc} S^2} = 
2 J_{44} - 4 J_{\perp} - 8\frac{J_{43} J_{43}'}{J_{33}} - 2 J_{33} - 2\frac{J_{43} J_{33}}{J_{43}'} -2 \frac{J_{43}' J_{33}}{J_{43}}\,.
\ee
\normalsize

\subsection{Collinear phases A}
Next we discuss the collinear state A that is mentioned in Fig.~\ref{fig:PD} of the main text. This state has ordering wavevector $\vec{Q}=(\pi,\pi,0)$ and is written as in Eq.~(\ref{fig:pipi0}) but with:
\small
\bea
\phi_5=\pi,~~ \phi_3=\phi_6+\pi,~~ \phi_6=\{0,\pi\} \,.
\eea
\normalsize
In this state all $J_{43}$ are satisfied, and only half of $J_{43}'$ are satisfied, and the two choices of $\phi_6$ fix which of the two halves are satisfied. The choice with $\phi_6\!=\!0$ is shown in Fig.~\ref{fig:Phases}\,(d).  
The energy corresponding to this solution is
\small
\bea
\frac{E_A}{N_{uc}S^2}=2J_{44}-4J_{\perp}-2J_{33}-8 J_{43} \,.
\eea
\normalsize
Finally, the expressions for the local fields are:
\small
\bea
\begin{array}{l}
h_1=h_2=(J_{44}-2 J_{43}\pm 2J_{43}'-J_{\perp})S, \\
h_3=h_4=(J_{44}-2J_{43} \mp 2J_{43}'-J_{\perp})S, \\
h_5=h_7=(-2J_{43}\mp 2J_{43}'-J_{33})S, \\
h_6=h_8=(-2J_{43}\pm 2J_{43}'-J_{33}) S, \\
h_9=h_{10}=-2J_{\perp} S\,,
\end{array}
\eea
\normalsize
where $\pm$ corresponds to $\phi_6=0$ or $\pi$. 
So here we have two types of Fe1 and two types of Fe2 spins.

\subsection{Collinear phases B}
Let us also discuss the collinear state B that is mentioned in Fig.~\ref{fig:PD} of the main text. This state has again ordering wavevector $\vec{Q}=(\pi,\pi,0)$ and is written as in Eq.~(\ref{fig:pipi0}) but with:
\small
\be
\phi_6=\pi,~~\phi_3=\phi_5=\{0,\pi\}\,. 
\ee
\normalsize
In this states all $J_{43}'$ are satisfied, and only half of $J_{43}$ are satisfied, and the choice of $\phi_5=\{0,\pi\}$ gives which of the two halves are satisfied. The choice with $\phi_5\!=\!0$ is shown in Fig.~\ref{fig:Phases}\,(e).  
The energy corresponding to this solution is
\small
\bea
\frac{E_B}{N_{uc}S^2}=2J_{44}-4J_{\perp}-2J_{33}-8 J_{43}' \,.
\eea
\normalsize
Finally, the expressions for the local fields are now:
\small
\bea
\begin{array}{l}
h_1=h_2=(J_{44}\pm 2J_{43}-2J_{43}'-J_{\perp}) S, \\
h_3=h_4=J_{44}\mp 2J_{43} - 2 J_{43}'-J_{\perp}) S, \\
h_5=h_7=2J_{43}-2J_{43}'-J_{33}) S, \\
h_6=h_8=\mp 2J_{43}-2J_{43}'-J_{33}) S, \\
h_9=h_{10}=-2J_{\perp}S\,,
\end{array}
\eea
\normalsize
where $\pm$ corresponds to $\phi_5=0$ or $\pi$. So we again have two types of Fe1 and two types of Fe2 sites.

\subsection{Boundaries between different phases}
Let us now establish the boundaries between the five phases discussed above. 
First of all, the orthogonal state is degenerate with the collinear phase A (B) when $J_{43}'=0$ ($J_{43}=0$). In these limits the classical degeneracy is actually much larger since we have decoupled chains, and each chain forms a N\'eel order along an independent spin direction. The collinear states A and B (as well as the orthogonal states A and B) are just two special members of this manifold. They are discussed here because they are the only collinear states in the phase diagram, together with the partly disordered, collinear phase that is stabilized by quantum fluctuations in the corner $J_{43}=J_{43}'=0$~\cite{rousochatzakis2012}.

Next, the boundary between the orthogonal and the mixed phase is determined by the line 
\small
\be
x_2= \frac{x_1}{\sqrt{4x_1^2-1}}
\ee 
\normalsize
where $x_1\equiv J_{43}/J_{33}$ and $x_2\equiv J_{43}'/J_{33}$. This is the brown line in Fig.~\ref{fig:PD} of the main text.
Next, the boundary between the mixed and the ferrimagnetic phase is determined by the line
\small
\be
x_2 = \frac{x_1}{2x_1-1}
\ee
\normalsize
which is where the two canting angles $\theta=\theta'=\pi/2$. This is the blue line in Fig.~\ref{fig:PD} of the main text.

\subsection{Generalized Luttinger-Tisza method of Lyons and Kaplan}
So far, we have only compared the energies of five possible configurations, but we have not proved which ones are the global ground states and where in the phase diagram. 
To this end, we will use Lyons and Kaplan's generalization~\cite{LyonsKaplan,*Kaplan} of the Luttinger-Tisza method~\cite{LT}.
We first rewrite the energy in momentum space
\small
\be
\mc{H}=\sum_{\vec{k}}\sum_{\nu,\mu=1}^{10} \Lambda_{\nu\mu}(\vec{k})  \vec{S}_{\vec{k},\nu}\cdot \vec{S}_{-\vec{k},\mu}
\ee
\normalsize
where $\vec{S}_{\vec{k},\nu}=\frac{1}{\sqrt{N_{uc}}} \sum_{\vec{k}} e^{-i \vec{k}\cdot\vec{r}} \vec{S}_{\vec{r},\nu}$, and the 10$\times$10 interaction matrix $\bs{\Lambda}$ is given by:
\small
\be
\bs{\Lambda}(\vec{k})=\frac{1}{2}\left(
\begin{array}{cccccccccc}
0 & J_{44} &0&0&J_{43}&J_{43}'&J_{43}&J_{43}'&J_{\perp}e^{-i\vec{k}\cdot\vec{c}}&0\\
J_{44}&0&0&0&J_{43}&J_{43}'&J_{43}&J_{43}'&J_{\perp}&0\\
0&0&0&J_{44}&J_{43}' e^{i\vec{k}\cdot\vec{t}_1}&J_{43}&J_{43}' e^{-i\vec{k}\cdot\vec{t}_2}&J_{43}e^{i\vec{k}\cdot(\vec{t}_1-\vec{t}_2)}&0&J_{\perp}e^{-i\vec{k}\cdot\vec{c}}\\
0&0&J_{44}&0&J_{43}' e^{i\vec{k}\cdot\vec{t}_1}&J_{43}&J_{43}' e^{-i\vec{k}\cdot\vec{t}_2}&J_{43}e^{i\vec{k}\cdot(\vec{t}_1-\vec{t}_2)}&0&J_{\perp}\\
J_{43}&J_{43}&J_{43}' e^{-i\vec{k}\cdot\vec{t}_1}&J_{43}' e^{-i\vec{k}\cdot\vec{t}_1}&0&0&J_{33} e^{-i\vec{k}\cdot\vec{t}_2}&0&0&0\\
J_{43}'&J_{43}'&J_{43}&J_{43}&0&0&0&J_{33} e^{i\vec{k}\cdot\vec{t}_1}&0&0\\
J_{43}&J_{43}&J_{43}' e^{i\vec{k}\cdot\vec{t}_2}&J_{43}' e^{i\vec{k}\cdot\vec{t}_2}&J_{33} e^{i\vec{k}\cdot\vec{t}_2}&0&0&0&0&0\\
J_{43}'&J_{43}'&J_{43}e^{-i\vec{k}\cdot(\vec{t}_1-\vec{t}_2)}&J_{43}e^{-i\vec{k}\cdot(\vec{t}_1-\vec{t}_2)}&0&J_{33} e^{-i\vec{k}\cdot\vec{t}_1}&0&0&0&0\\
J_{\perp}e^{i\vec{k}\cdot\vec{c}}&J_{\perp}&0&0&0&0&0&0&0&0\\
0&0&J_{\perp}e^{i\vec{k}\cdot\vec{c}}&J_{\perp}&0&0&0&0&0&0
\end{array}
\right)\,.\nonumber
\ee
\normalsize
Next we replace the strong constraints $\vec{S}_{\vec{r,\nu}}^2=S^2$, with the `generalized weak constraint' of Lyons and Kaplan~\cite{LyonsKaplan,*Kaplan},
\small
\be 
\sum_{\vec{r},\nu} \alpha_\nu \vec{S}_{\vec{r,\nu}}^2=S^2 N_{uc}\sum_\nu \alpha_\nu\,,
\ee
\normalsize
where the set $\{\alpha_\nu\}$ is a {\bf fixed} set of numbers. 
Minimizing the associated Langrange multiplier problem leads to the equations:
\small
\bea
\sum_{\nu} \Lambda_{\mu\nu}(-\vec{q}) \vec{S}_{\vec{q},\nu}=\lambda \alpha_\mu \vec{S}_{\vec{q},\mu} \Rightarrow 
\sum_{\nu} \widetilde{\Lambda}_{\mu\nu}(-\vec{q}) \widetilde{\vec{S}}_{\vec{q},\nu}=\lambda \widetilde{\vec{S}}_{\vec{q},\mu}\,,
\label{eq:eval}
\eea
\normalsize
where $\lambda$ is the Langrange multiplier,  
\small
\be\label{eq:Ltilde}
\widetilde{\Lambda}_{\mu\nu}\!=\!\frac{\Lambda_{\mu\nu}}{\sqrt{\alpha_\mu\alpha_\nu}},~~
\widetilde{\vec{S}}_{\vec{k},\mu}\!=\!\sqrt{\alpha_\mu}\,\vec{S}_{\vec{r},\mu}\,.
\ee
\normalsize 
The total ground state energy can be expressed as 
\be 
E=N_{uc}S^2\lambda \sum_{\nu=1}^{10} \alpha_\nu\,,
\ee 
and since the set of $\{\alpha_\mu\}$ is fixed, the minimum energy of this `generalized soft constraint problem' is obtained by minimizing $\lambda$, i.e. by choosing the minimum eigenvalue of the matrix $\widetilde{\bs{\Lambda}}$ over the whole Brilouin zone. 
Now the key point is the following~\cite{LyonsKaplan,*Kaplan}: All solutions of the original `hard constraint problem' satisfy the `generalized soft constraint'. So all solutions of the former problem are enclosed within the domain of solutions of the latter problem. So if we find the {\it minimum} solution over the latter domain, then this must be the minimum solution of the `hard constraint problem' as well. 

The main problem then reduces to `guessing' the right set $\{\alpha_\mu\}$. 
To this end we go back and identify 
\be
\lambda\alpha_\mu \to h_\mu/(2S),
\ee 
where $h_\mu$ is the strength of the local mean field at site $\mu$. And for the latter we can just try out the local fields from some of the above five configurations. Accordingly, we have checked numerically the eigenspectrum of the matrices $\widetilde{\bs{\Lambda}}_{\text{ortho}}$ and $\widetilde{\bs{\Lambda}}_{\text{ferri}}$, obtained from Eq.~(\ref{eq:Ltilde}) with the local fields associated with the orthogonal and the ferrimagnetic state, respectively.  
We find the following numerically:
\\
\begin{enumerate}

\item Within the `orthogonal' region of Fig.~\ref{fig:PD} of the main text, the minimum eigenvalue of the matrix $\widetilde{\Lambda}_{\text{ortho}}$, sits at the wavevector $\vec{Q}=(\pi,\pi,0)$, it is two-fold degenerate, and the associated eigenvectors can be combined to give precisely the two orthogonal states. This proves that the two orthogonal states are the absolute minima of the energy in this region of the phase diagram. 

\item Within the `ferrimagnetic' region of Fig.~\ref{fig:PD} of the main text, the minimum eigenvalue of the matrix $\widetilde{\Lambda}_{\text{ferri}}$, sits at the wavevector $\vec{Q}=(0,0,0)$, it is not degenerate, and the associated eigenvector gives precisely the ferrimagnetic state. This proves that the ferrimagnetic state is the absolute minimum of the energy in this region of the phase diagram. 

\item Within the `mixed' region of Fig.~\ref{fig:PD} of the main text, the minimum eigenvalue of the matrix $\widetilde{\Lambda}_{\text{ferri}}$, sits at the wavevector $\vec{Q}=(\pi,\pi,0)$, and it is two-fold degenerate, and the two solutions correspond to the orthogonal states. Likewise the minimum eigenvalue of the matrix $\widetilde{\Lambda}_{\text{ortho}}$, sits at the wavevector $\vec{Q}=(0,0,0)$, it is not degenerate and corresponds to the ferrimagnetic solution. 
This proves that neither the ferrimagnetic nor the orthogonal states are the absolute minima in this region of the phase diagram, and reveals that the absolute minimum should be searched in the form of a linear combination of the two states instead. 
\end{enumerate}
Altogether, we have proved that the phases shown in Fig.~\ref{fig:PD} of the main text correspond to the absolute minima configurations.

\section{Qualitative behavior of the local spin lengths in the isotropic spin model}
Here we would like to briefly show the qualitative difference between the $T$-dependence of the interlayer Fe3 spins with that of the Fe1 and Fe2 sites. 
Figure~\ref{fig:SL} shows the $T$-dependence of the local spin lengths as obtained from a hybrid, quantum-mechanical mean-field approach. In this approach, the Fe1 and Fe3 sites are treated within single-site quantum mean-field theory, while the Fe2 spins are treated in pairs, by solving the problem of spin-5/2 dimers in some external self-consistent mean field. 
In this calculation we have used the exchange parameters reported in the main text.

As expected, the results overestimate the ordering temperature (which is in fact also determined by $J_\perp$, see e.g. Ref.~[\onlinecite{CuP2O6}]), but the qualitative behavior for the spin lengths is already captured by this mean-field approach. 
Specifically, we see that the Fe3 moments grow far more slowly with decreasing $T$ below the ordering temperature. This is a key aspect for the reorientation transition discussed in the main text. 

Another noteworthy feature is the value of the Fe2 moments at zero temperature, which unlike the Fe1 spins, does not go to the maximum possible value of $5/2$, but is slightly reduced. This reduction is related to the fact that the ground state wavefunction of the dimer mean-field problem has a finite singlet component due to the finite $J_{33}$ coupling. Specifically, the local exchange field exerted on the two sites of the Fe2 dimer has a large staggered component (it has no uniform component at all when $J_{43}=J_{43}'$, see Ref.~[\onlinecite{rousochatzakis2012}]). As a result, the ground state wavefunction of the dimer mean-field problem is a combination of the singlet (as in the absence of exchange field), the triplet, etc (up to $S=5$). This admixture is responsible for the finite staggered polarization on the dimer.

For comparison we also show the results from a quantum, single-site mean-field theory on all sites, where all spin lengths go to $5/2$ at zero temperature.

\begin{figure}
\includegraphics[width=0.48\textwidth,angle=0,clip=true,trim=0 0 0 0]{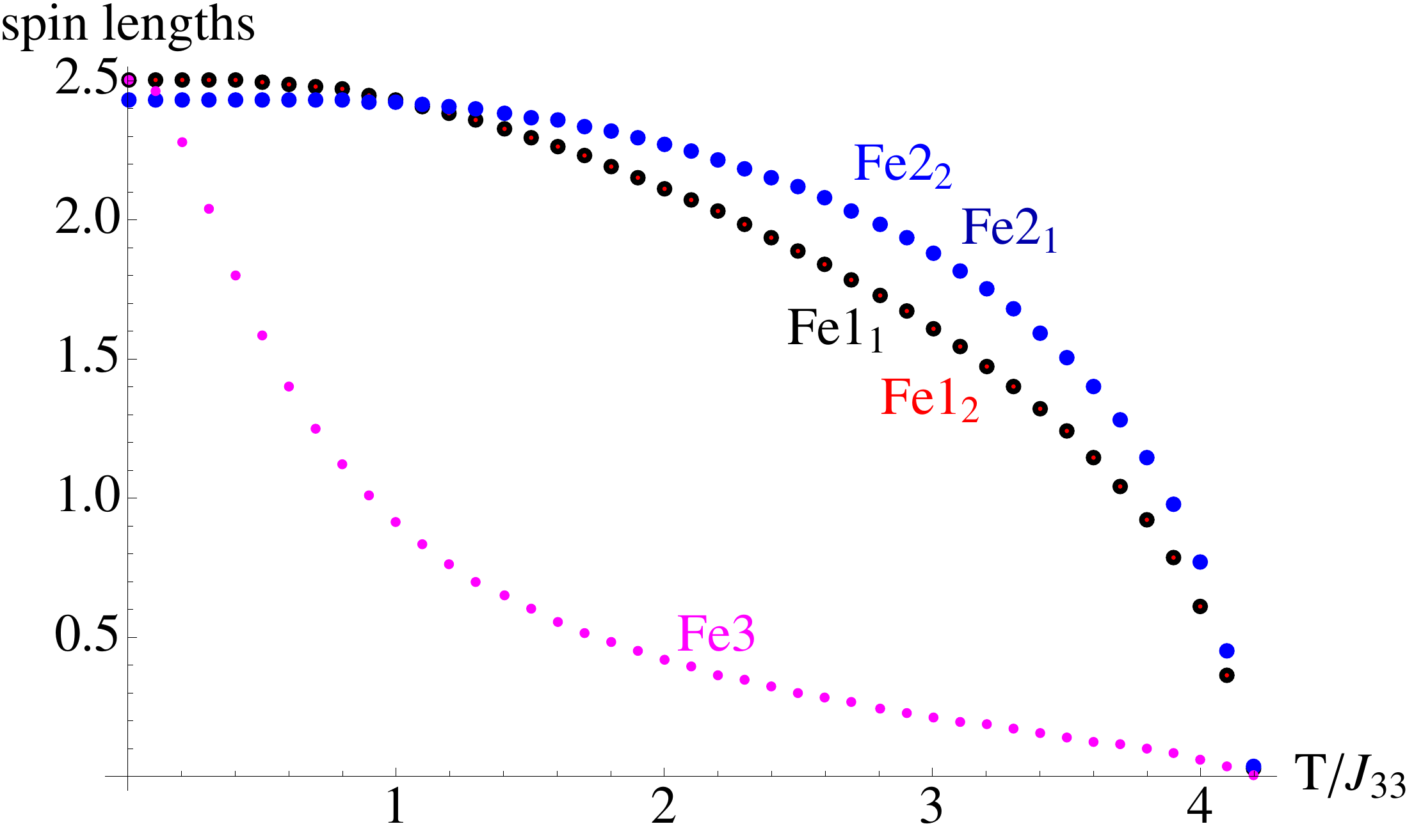}~
\includegraphics[width=0.48\textwidth,angle=0,clip=true,trim=0 0 0 0]{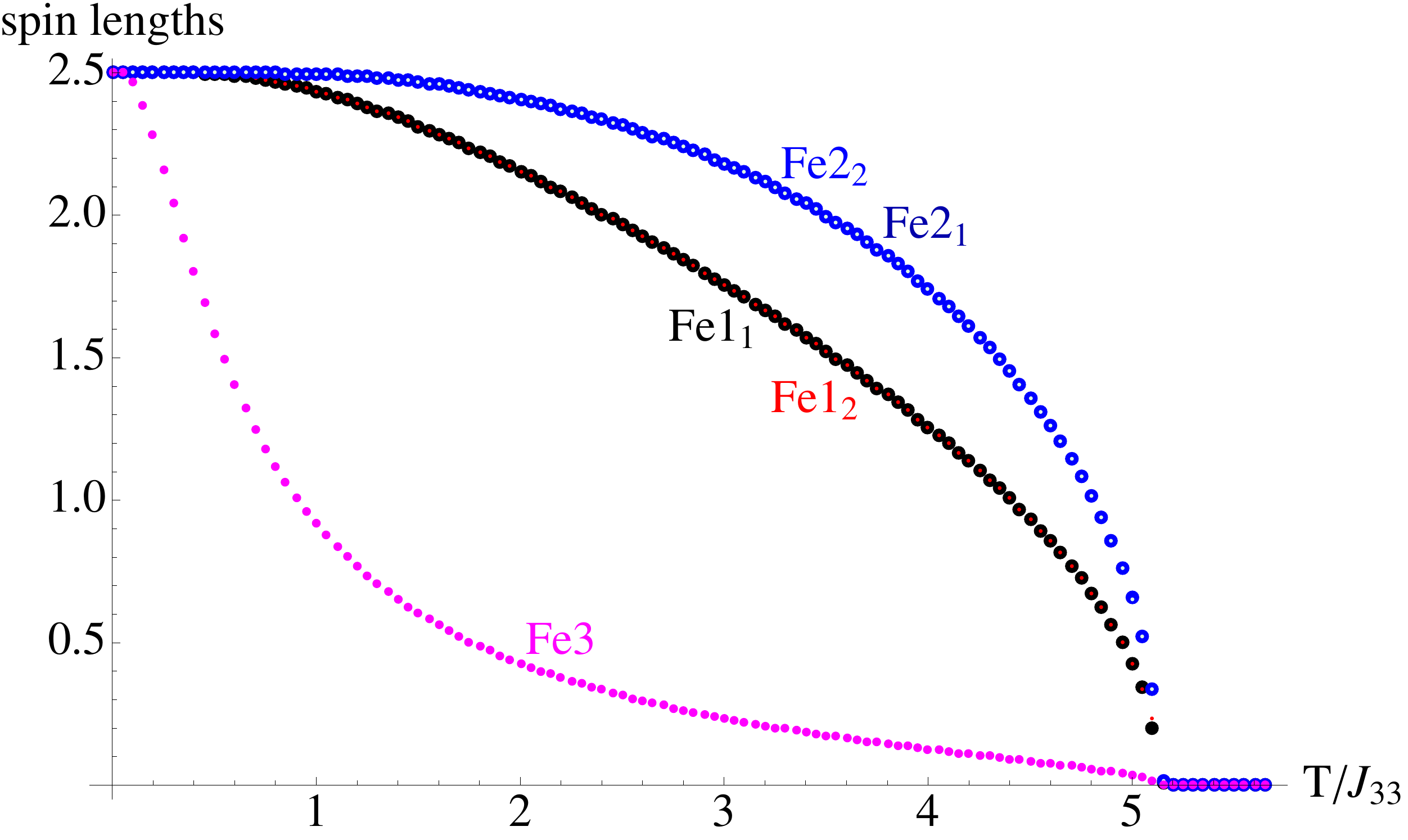}
\caption{\label{fig:SL} $T$-dependence of the local moments as obtained from a hybrid self-consistent quantum mean-field theory (left) and a single-site quantum mean-field theory (right). The exchange parameters used are the ones reported in the main text.}
\end{figure}


\end{widetext}

\end{document}